\documentclass{article}

\usepackage[preprint]{neurips_2021}

\usepackage{graphicx} %Loading the package
\graphicspath{{FIG/}} %Setting the graphicspath

\usepackage[utf8]{inputenc} % allow utf-8 input
\usepackage[T1]{fontenc}    % use 8-bit T1 fonts
\usepackage{hyperref}       % hyperlinks
\usepackage{url}            % simple URL typesetting
\usepackage{booktabs}       % professional-quality tables
\usepackage{amsfonts}       % blackboard math symbols
\usepackage{nicefrac}       % compact symbols for 1/2, etc.
\usepackage{microtype}      % microtypography
\usepackage{xcolor}         % colors

\usepackage{amsmath}

\author{
  Joshua Pan$^1$   \And
  Jason J. Kwon$^1$   \And
  Jessica A. Talamas$^1$ 
  \And
  Ashir A. Borah$^1$
  \And
  Francisca Vazquez$^1$
  \And
  Jesse S. Boehm$^1$
  \And
  Aviad Tsherniak$^1$
  \AND
  Marinka Zitnik$^2$
  \And
  James M. McFarland$^1$
  \And
  William C. Hahn$^1$
  \AND
  \textmd{\(^1\) Broad Institute \(^2\) Harvard Department of Biomedical Informatics}\And
  \textmd{Correspondence}: \texttt{william\_hahn@dfci.havard.edu}
}
\newcommand{\xhdr}[1]{\vspace{1.3mm}\noindent{{\bf #1.}}}

\title{Sparse dictionary learning recovers pleiotropy from human cell fitness screens}
\begin{document}
\maketitle
\begin{abstract}
In high-throughput functional genomic screens, each gene product is commonly assumed to exhibit a singular biological function within a defined protein complex or pathway. In practice, a single gene perturbation may induce multiple cascading functional outcomes, a genetic principle known as pleiotropy. Here, we model pleiotropy in fitness screen collections by representing each gene perturbation as the sum of multiple perturbations of biological functions, each harboring independent fitness effects inferred empirically from the data. Our approach (‘Webster’) recovered pleiotropic functions for DNA damage proteins from genotoxic fitness screens, untangled distinct signaling pathways upstream of shared effector proteins from cancer cell fitness screens, and learned aspects of the cellular hierarchy in an unsupervised manner. Modeling compound sensitivity profiles in terms of genetically defined functions recovered compound mechanisms of action. Our approach establishes a sparse approximation mechanism for unraveling complex genetic architectures underlying high-dimensional gene perturbation readouts.

\end{abstract}

\section{Introduction}

CRISPR-Cas9 technology has made gene perturbation a routine practice, and genome-scale screens can be readily performed across diverse cell contexts to measure physiological outcomes such as cell fitness. After collecting such datasets, a key challenge is to infer the genetic architecture underlying the observed fitness effects, such that individual genes become mapped to putative biological functions essential in specific cell contexts for cell fitness.

A foundational aspect of genetic architecture is pleiotropy, which states that gene products can participate in multiple independent biological functions. Although pleiotropy is pervasive, our ability to account for pleiotropy within collections of cell fitness screens is limited. Because each gene perturbation is measured across many cell contexts, to capture pleiotropy one must first describe a set of biological functions that vary independently across cell contexts, then define a one-to-many mapping of genes to these functions. Both of these steps are challenging to perform in high-dimensional data, and the lack of a unified conceptual framework results in different calculations of pleiotropy which cannot be directly compared (Costanzo et al., 2016; Dudley et al., 2005; Koch et al., 2017). 

\xhdr{Present work}
Here, we propose a framework that exploits pleiotropy to structure latent representations of biological function learned from fitness data. Our approach (‘Webster’) takes a large gene perturbation matrix as input and infers a set of biological functions, which we refer to as a dictionary, such that each gene perturbation can be approximated as a combination of a small number of these dictionary elements. Regularizing the dictionary using the gene co-fitness graph results in individual dictionary elements mapping to interpretable biological modules. By applying Webster to CRISPR-Cas9 fitness screen collections performed in human cells under a variety of conditions, we explored the layers of functional impact resulting from single gene perturbation; jointly embedded gene and functional dependencies to chart fitness landscapes; and projected new perturbations into the learned reference space.

\section{Methods}
\subsection{Overview}
\begin{figure}[h]
\centering
\includegraphics[width=\textwidth]{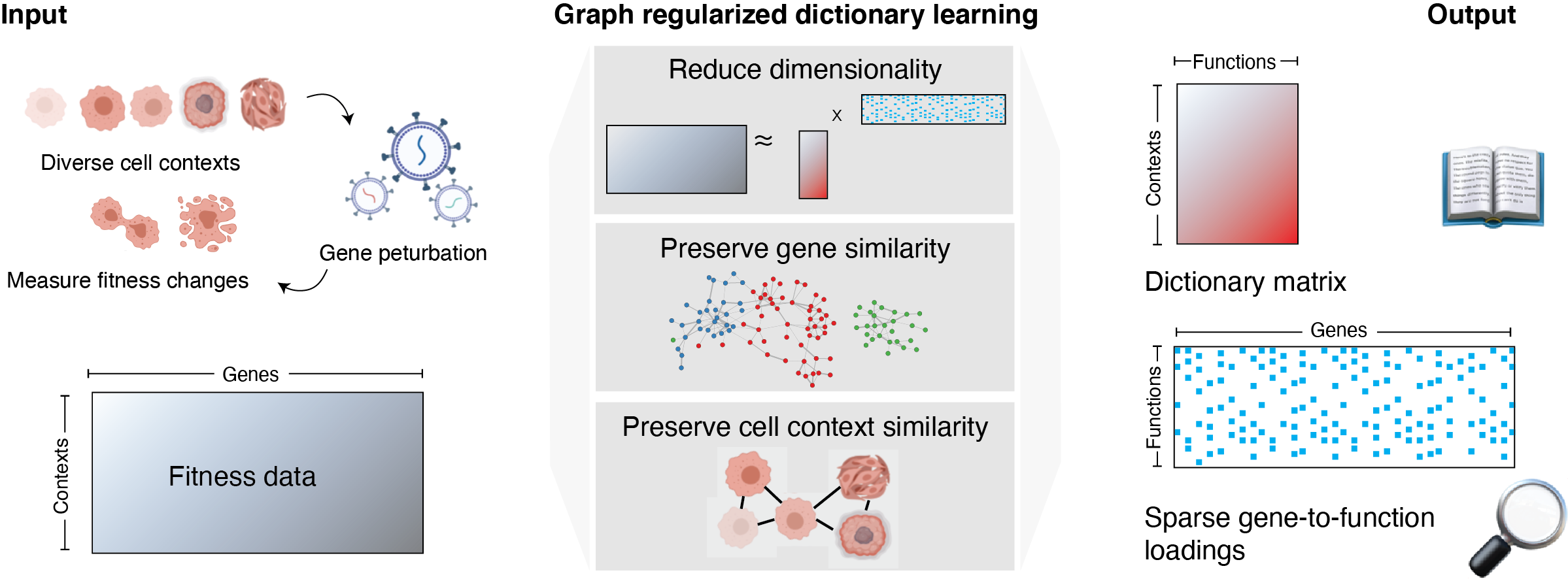}
\caption{\textit{Pleiotropy underlying the fitness effects of gene perturbation can be approximated using dictionary learning.} Fitness screen collections measure changes in cell growth rate following gene perturbation across diverse cell contexts. Webster applies graph-regularized dictionary learning to this data to discover latent variables corresponding to biological functions. Webster returns: 1) A dictionary matrix containing the fitness effect of perturbing each inferred biological function and 2) sparse gene-to-function loadings. Using this information, each measured gene effect can be approximated as a sparse linear combination of these latent functional effects, scaled by the appropriate loadings. Given the number of latent functions ($k$) and a hard sparsity parameter ($t$), Webster minimizes the total approximation error while preserving the local structure of genes and cell contexts in its reduced-dimensional representations.}
\label{f1}
\end{figure}

Given a set of gene perturbation measurements, we wish to infer a dictionary of latent variable elements, such that the effects of each gene perturbation can be approximated as a mixture of dictionary elements. Furthermore, dictionary elements should correspond to interpretable biological functions learned empirically from the data with no outside knowledge. To perform this inference, we employed dictionary learning via Webster (Figure $\ref{f1}$). Webster receives as input an $n \times m$ matrix of fitness effects, where $n$ is the number of cell contexts, and $m$ is the number of genes. (A fitness effect is the change in cell number upon gene perturbation.) 

Webster models the fitness effect matrix in terms of $k$ latent biological functions learned from the data, with $k<m$ and $n$. The output of Webster is two low-rank matrices: (1) an $n \times k$ dictionary matrix capturing the fitness effect of losing one of $k$ inferred functions across $n$ cell contexts, and (2) a $k \times m$ loadings matrix encoding the sparse approximation of each of $m$ gene effects in terms of $t$ dictionary elements, with $t << k$. Formally, this approach corresponds to an approximation task called “sparse dictionary learning” (Rubinstein et al., 2010), with connections to sparse matrix factorization and dimensionality reduction (Cleary et al., 2017; Kim and Park, 2007; Stein-O’Brien et al., 2018).

Webster incorporates several biologically-inspired priors to constrain its dictionary learning (Figure $\ref{f1}$, Figure $\ref{sf1}$). First, cell contexts that depend on similar genes for proliferation are encouraged to depend on similar functions (as represented in the dictionary matrix). Second, genes with similar perturbation outcomes in the original data are encouraged to be loaded onto similar sets of functions, capturing modular pleiotropy (Wang et al., 2010) between co-essential pathways (Wainberg et al. 2021). Finally, $t$ is chosen to be small such that each gene effect is modeled in terms of a few functions, reflecting the observed sparsity across genotype-phenotype maps in model organisms (Wang et al., 2010). Latent variables obeying these locality and sparsity constraints will capture coherent gene programs that are selectively dependent in particular cell contexts, thereby informing biological function (Keeling et al., 2019; Wagner and Zhang, 2011). 

\subsection{Model description}
Webster encompasses three steps: (1) preprocessing raw fitness data; (2) dictionary initialization with $k$-medoids; (3) graph-regularized dictionary learning. Preprocessing removes low-variance gene effects and corrects batch effects between cell contexts prior to dictionary learning. Then, $k$-medoids defines an initial $n \times k$ dictionary that clusters the data. From this starting point, dictionary learning is performed using an objective function balancing (i) approximation error, (ii) smoothness over a nearest-neighbor graph of genes and (iii) smoothness over a nearest-neighbor graph of cell contexts. This is performed via dual graph-regularized $k$-SVD (Yankelevsky and Elad, 2016, 2020), which optimizes $k$ overlapping subspaces of genes and takes the first eigenvector of the subspace as its representative dictionary element. Each gene effect is then linearly approximated using $t$ dictionary elements via orthogonal matching pursuit (Pati et al., 1993), thereby capturing statistically independent components of variance. A detailed methodological overview is in the \hyperref[appendix]{Appendix}.

\subsection{Implementation and data sources}
We applied Webster to published CRISPR-Cas9 fitness screens in a human cell line exposed to a diverse set of genotoxic stressors  (Olivieri et al., 2020). Preprocessing the 31 genome-scale screens yielded 304 high variance fitness genes (Figure $\ref{sf2}$A). After a hyperparameter sweep (Figure $\ref{sf2}$B), Webster successfully reconstructed the original gene perturbation data using an inferred dictionary composed of ten elements, using two dictionary elements to represent each gene perturbation (Table S1). We matched each dictionary element to a biological function by annotating its most sensitive genotoxic stressors and its top loaded genes, using literature resources that were hidden during model training.

Next, we scaled Webster to a $\sim$ 20 times larger dataset of 675 cancer cell line fitness screens, the Cancer Dependency Map (Meyers et al., 2017). We preprocessed the data to include 2,921 gene effects exhibiting high variance across cell lines. After a hyperparameter sweep (Figure $\ref{sf3}$A-B), we recovered 220 dictionary elements and approximated each gene effect using up to four dictionary elements (Table S2). We further validated that Webster was reproducible across random runs over identical parameters (Figure $\ref{sf3}$C), robust to noise (Figure $\ref{sf3}$D), and that dictionaries trained on one cancer fitness dataset could successfully model perturbations from another dataset performed with a different CRISPR-Cas9 guide library and cell culture conditions (Behan et al., 2019) (Figure $\ref{sf3}$E). We matched each dictionary element to a biological function by annotating the strongest loaded genes, supported by consensus cell line features that explain the fitness effect captured by the element. Nine out of ten dictionary elements captured the fitness effect of losing literature-supported functions, whereas the remaining elements captured technical factors such as common essential effects (Table S2).

\section{Results}
First, we benchmark Webster's inferences against other common latent variable methods. Then, we explore biological insights that emerge from Webster factorizations. 

\subsection{Benchmarking}
We compared Webster's inferences against three commonly used latent variable methods: Principal Components Analysis (PCA), Independent Components Analysis (ICA), and $k$-medoids (Table 1). PCA and ICA learned latent variables that well approximated the data but lacked sparsity and interpretability. By relaxing the one-hot sparsity assumption in $k$-medoids to allow pleiotropy, Webster improved the recovery of ground-truth genesets as well as reduced the approximation error of the overall factorization.

\begin{table*}[h]
\caption{\textit{Benchmarking results for Webster $vs.$ common latent variable methods.}  The mean over 5 randomized runs is reported for: root mean square error; the smoothness over the gene co-fitness graph, calculated as the mean squared difference between gene loadings for gene pairs strongly correlated in the original data; sparsity of gene loadings; entropy and purity of the factorizations against ground truth gene annotations, taken from (Olivieri et al., 2020) and (Sanchez-Vega et al., 2018). Entropy and purity calculated as in (Kim and Park, 2007). Best scores are reported in bold. $k = 10$ for the genotoxic dataset, and 220 for DepMap.}
\label{tab:knn}
\begin{center}
\begin{small}
\begin{sc}
\begin{tabular}{ccccccc}
\toprule
$Dataset$ & $Method$ & RMSE & \textsc{Smoothness} & \textsc{Sparsity} & Entropy & Purity  \\
\midrule
\text{Genotoxic } & PCA &1.88& 7.49 & 0.300 & 0.649& 0.416 \\[2.5pt]
& ICA  &2.59& \textbf{4.49} & 0.383 & 0.695&0.452\\[2.5pt]
& K-Medoids &2.20& 7.82& \textbf{1.000} & 0.484 & 0.602\\[2.5pt]
& Webster  & \textbf{1.80} & 4.64 & 0.889 & \textbf{0.462} & \textbf{0.624}\\[2.5pt]

\hline\\[2.5pt]
\text{DepMap} & PCA&0.782&3.3e2&0.299&0.385&0.607\\[2.5pt]
&ICA&\textbf{0.774}&2.7e3&0.262&0.669&0.385\\[2.5pt]
&K-Medoids&1.88&4.8e2&\textbf{1.000}&0.234&0.689\\[2.5pt]
&Webster&1.17&\textbf{2.1e2}&0.949&\textbf{0.101}&\textbf{0.882}
\end{tabular}
\end{sc}
\end{small}
\end{center}
\end{table*}

\subsection{Learning representations of DNA damage functions from genotoxic fitness screens}

Next, we explored the biological insights recovered by Webster from the genotoxic screening data (Figure $\ref{f2}$A). As an example, Webster’s first dictionary element (Figure $\ref{f2}$B) captured the negative fitness effects of DNA adduct-inducing agents (UV light, illudin S, and BPDE) (Figure $\ref{f2}$C), which are toxic in the absence of nucleotide excision repair (NER). The top four genes loaded on this dictionary element were classical NER members (ERCC8, GTF2H5, UVSSA, ERCC5), and the fifth (STK19) was a recently discovered pathway member (Boeing et al., 2016; Olivieri et al., 2020). Indeed, the strength of a gene’s loading on this element was a sensitive and specific predictor of NER pathway membership (Figure $\ref{sf2}$C, AUROC = 0.9).

We identified other dictionary elements that captured the effects of DNA damage pathways (Figure $\ref{f2}$D), with canonical pathway members loaded onto the appropriate dictionary elements (Figure 2D). PCA and ICA failed to separate these independent pathways into individual components (Figure $\ref{sf2}$C, Table 1). From the fitness data alone and without geneset information as input, Webster learned a set of maximally informative dictionary elements capturing the functional effects of losing biological pathways that respond genotoxic stress.

\subsection{Pleiotropy underlies the DNA damage response}

\begin{figure}
\centering
\includegraphics[width=\textwidth]{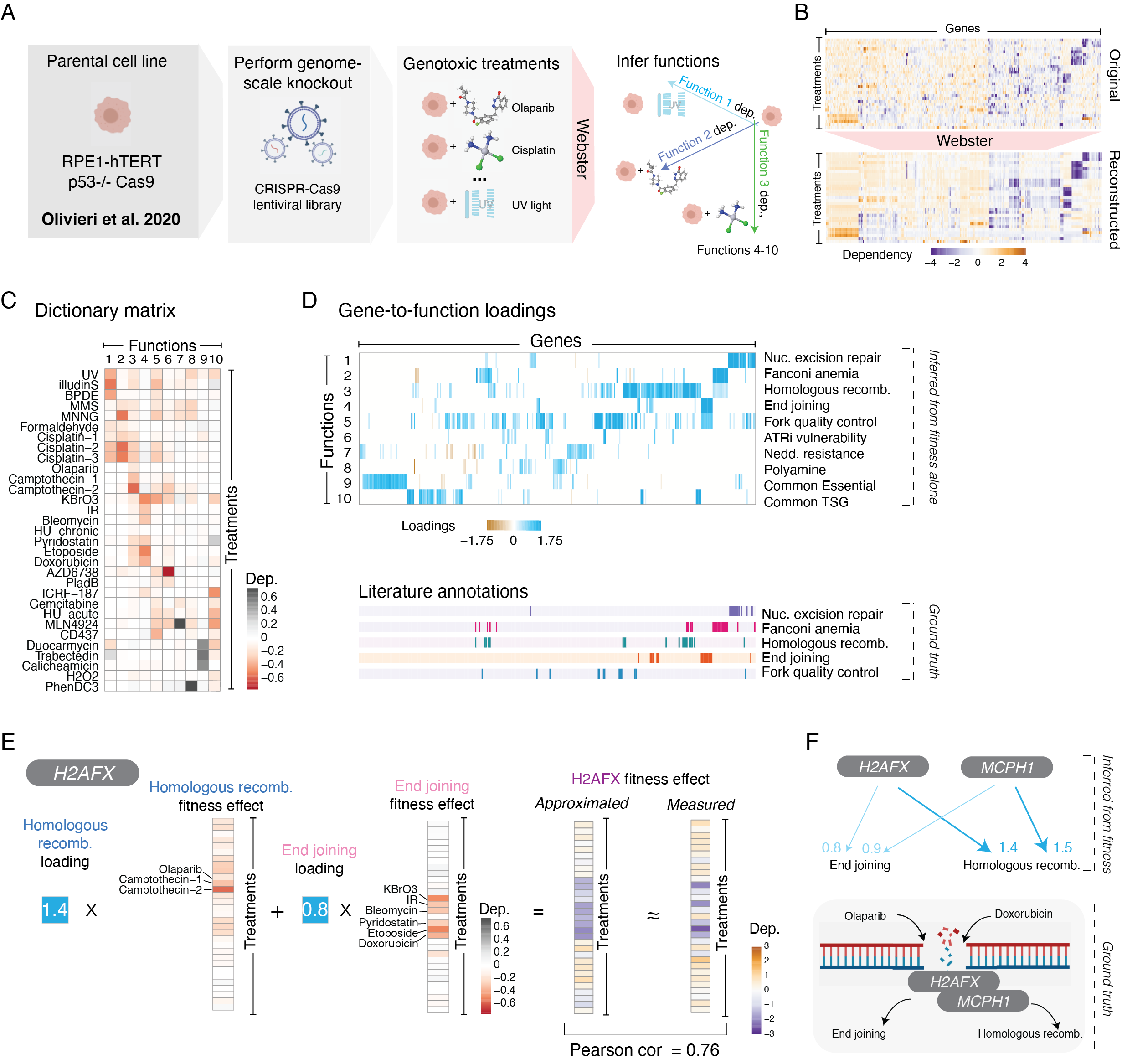}
\caption{\textit{Pleiotropy underlies the DNA damage response to genotoxins in a human cell line.} A.	The immortalized human cell line RPE1-hTERT harboring a genome-scale CRISPR-Cas9 knockout library was subjected to 31 genotoxic stressors at a sub-lethal dose, resulting in a genotoxic fitness screen collection (Olivieri et al., 2020). From this data matrix, Webster was parameterized to infer ten biological functions and approximate each gene effect as a sparse mixture of two functional effects.
B.	Top: The original fitness data, preprocessed to a set of 304 high-variance fitness gene effects from 31 treatment conditions, shown as a hierarchically clustered heatmap. Bottom: Webster’s approximation of the data, with each gene effect approximated as a sparse mixture of two inferred functions. The order of genes and treatments is preserved between panels.
C.	The dictionary matrix. Each column of the dictionary captures the inferred fitness effect of depleting an biological function learned from the data.
D.	The loadings matrix. Top: Sparse gene-to-function loadings for the 304 fitness genes. Each gene (column) has two non-zero loadings, encoding the model’s sparse representation of its gene effect. Bottom: Literature curated gene annotations, defined by (Olivieri et al., 2020). Gene order is preserved between panels. TSG = Tumor Suppressor Gene.
E.	Gene effect decomposition. Webster decomposes H2AFX knockout as a mixture of two functional effects related to DNA double stranded breaks. The first function, Homologous Recombination, has a fitness effect induced by olaparib and camptothecin, etc. The second, End Joining, has a fitness effect induced by doxorubicin and etoposide, etc. Webster faithfully modeled the H2AFX gene effect as the sum of these two functional effects, scaled by their respective loadings (Pearson  = 0.76).
F.	Top: Relationships learned from fitness data for H2AFX and MCPH1, an obligate H2AFX interactor. Each arrow corresponds to an inferred gene-to-function loading. Bottom: Illustration of H2AFX/MCPH1’s shared roles as DNA double stranded break sensors upstream of Homologous Recombination and End Joining.}
\label{f2}
\end{figure}

\begin{figure}
\centering
\includegraphics[width=\textwidth]{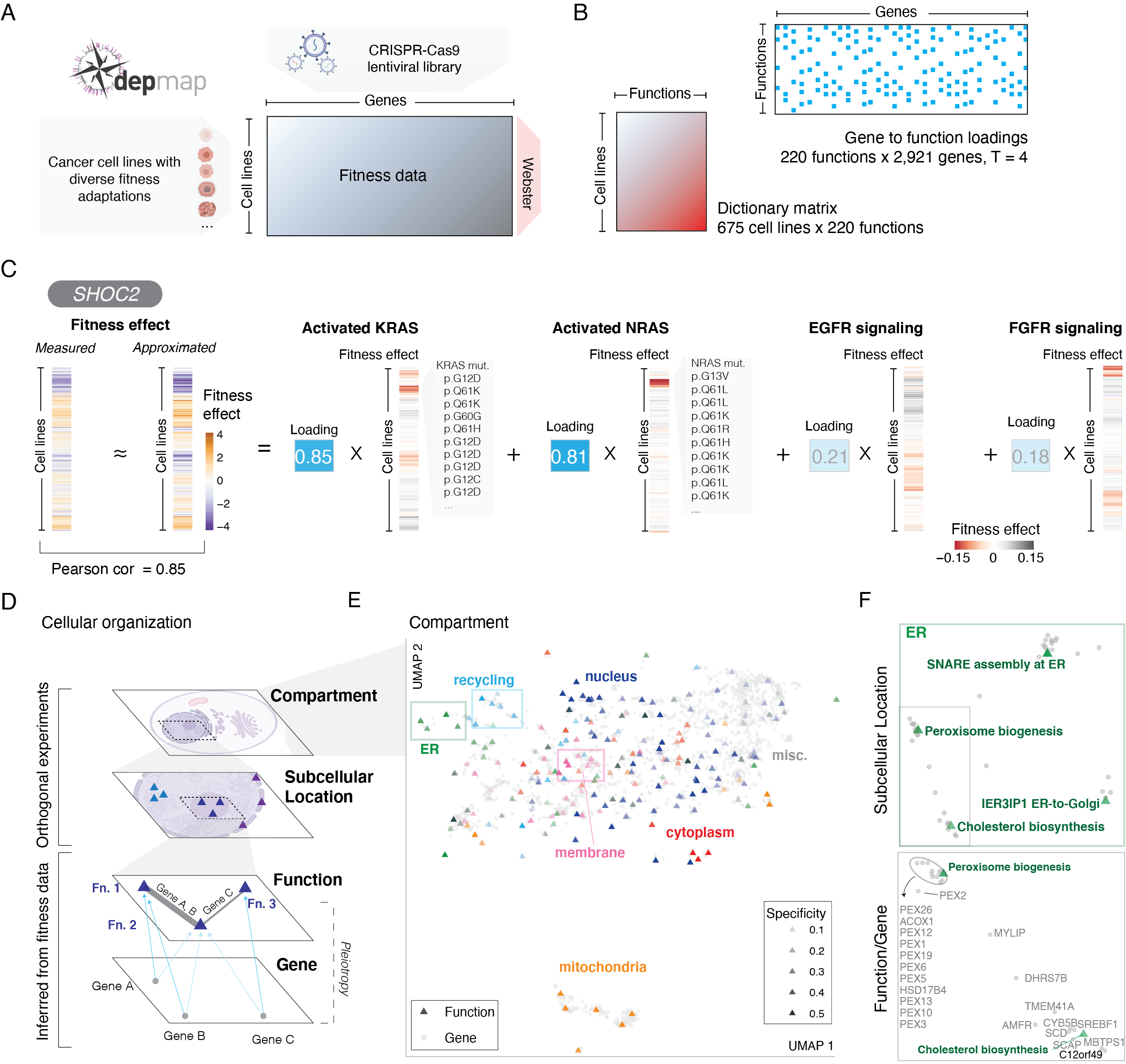}
\caption{\textit{Pleiotropy underlies the fitness effect of gene knockout across human cancer cell lines.} A.	The Cancer Dependency Map (DepMap) fitness screen collection. Individual human cancer cell lines were screened for genetic dependencies for cell growth by comparing cell counts before and after infection with a genome-scale CRISPR-Cas9 gene knockout library.
B.	The DepMap data was preprocessed to a set of 2,921 high variance fitness genes screened across 675 cell lines. From this data alone, Webster learned a dictionary matrix of 220 fitness effects reflecting inferred biological functions, and approximated each gene effect in terms of four functional effects.
C.	Webster approximated the fitness effect of SHOC2 knockout as a mixture of four functional effects (Activated KRAS, Activated NRAS, EGFR signaling and FGFR signaling), each of which were strongest in cell lines harboring corresponding genomic alterations (KRAS mutation, NRAS mutation, activated EGFR and FGFR expression, respectively). This decomposition reflects the pleiotropic interactions underlying SHOC2’s overall function downstream of these signaling pathways. 
D.	Conceptual overview of cellular organization. After learning gene functions with Webster from cancer fitness data alone, experimentally derived subcellular localization data (Go et al., 2021) were used to annotate Webster’s functions across 20 subcellular locations within a total of 7 cellular compartments. 
E.	Joint embedding of fitness effects for genes and functions inferred from DepMap data, with functions colored by their specificity for one of seven cellular compartments -- mitochondria, endoplasmic reticulum (ER), recycling, membrane, nucleus, cytosol and miscellaneous. Subcellular location data was not used during the training of the Webster model.
F.	Top: Insets from E. detailing functions within the ER, capturing specific subcellular locations or protein complexes within these broad compartments. Bottom: Insets from above panel detailing genes embedded nearby their pleiotropic gene functions.}
\label{f3}
\end{figure}

We next examined how learned mixtures of functional effects approximated pleiotropic gene effects. The H2AFX histone protein is phosphorylated in response to DNA double stranded breaks. Webster approximated the effect of perturbing H2AFX as a mixture of two functional effects scaled by their loadings (Pearson correlation = 0.76, Figure $\ref{f2}$E): 

\makebox[\linewidth]{$\textit{H2AFX} \approx 1.4 * \textit{Homologous Recombination} + 0.8 * \textit{End Joining}$}

MCPH1 is an obligate H2AFX interactor at double stranded break sites. Webster approximated the effect of depleting MCPH1 using identical functional effects and similar loadings (Figure $\ref{f2}$F). Homologous Recombination and End Joining both repair double stranded breaks, but are preferentially activated under different conditions (Figure $\ref{f2}$E and F). By constraining Webster to approximate a large number of gene perturbations with a small number of dictionary elements, it parsimoniously modeled perturbations of double stranded break sensors as mixtures of the underlying double stranded break repair pathways.

We note that linear algebra operations in our interpretable latent space reflect the underlying structure of pleiotropic gene relationships. RAD51B is a Homologous Recombination gene as well as risk allele for Fanconi Anemia. In our model, $\textit{H2AFX}  - \textit{End Joining} + \textit{Fanconi Anemia} \approx \textit{RAD51B}$ (Pearson cor = 0.69). This is analogous to linear semantics underlying word embeddings in which $\textit{king} - \textit{man} + \textit{woman} \approx \textit{queen}$ (Mikolov et al., 2013).

\subsection{Learning pleiotropic biological functions from cancer cell line fitness screens}

From the larger Cancer Dependency Map dataset, Webster modeled 2,921 gene effects as pleiotropic mixtures of  220 functional effects learned empirically from cancer fitness data (Figure $\ref{f3}$A-B). For instance, SHOC2 is part of the oncogenic RAS signaling pathway. Webster approximated the effect of SHOC2 depletion in 675 cell lines in terms of four functional effects from the dictionary (Pearson correlation = 0.85, Figure $\ref{f3}$C):  

\makebox[\linewidth]{$\textit{SHOC2} \approx  0.9 * \textit{Activated KRAS} + 0.8 * \textit{Activated NRAS} +$}
\makebox[\linewidth]{$  0.2 * \textit{EGFR Signaling} + 0.2 * \textit{FGFR Signaling}$}

Cell lines with the strongest fitness effects per function exhibited matched molecular alterations: KRAS hotspot mutations, NRAS hotspot mutations, activated EGFR and high FGF expression, respectively (Figure 3C, Figure $\ref{sf3}$F). These genomics features were unused during model training. By looking globally across fitness data alone, Webster discovered cell contexts with distinct activated signaling pathways that influence SHOC2 function, thereby revealing functional components underlying a single gene perturbation. 

A biochemical basis for pleiotropy arises when protein complexes share common subunits but use them to perform distinct functions. From the fitness data alone, Webster learned independent representations for STAGA and ATAC complexes. From these, the effect of perturbing a shared subunit was represented as $KAT2A \approx 1 * \textit{STAGA complex} + 0.2 * \textit{ATAC complex} + \cdots $, while the effects of perturbing complex-specific subunits were represented as  $ADA2A \approx \textbf{0} * \textit{STAGA complex} + 0.26 * \textit{ATAC complex} + \cdots$ and $ADA2B \approx 2.2 * \textit{STAGA complex} + \textbf{0} * \textit{ATAC complex} + \cdots$. Other modular complexes such as SWI/SNF, Mediator, and Integrator exhibited similar patterns. Overall, 53\% of the gene effects that were well approximated by Webster (Pearson cor $\geq$ 0.4) exhibited evidence of pleiotropy (loadings $\geq$ 0.25 on at least two functions), placing cancer cells in line with other organisms exhibiting majority pleiotropic genes (Wang et al., 2010).

\begin{figure}
\centering
\includegraphics[width=\textwidth]{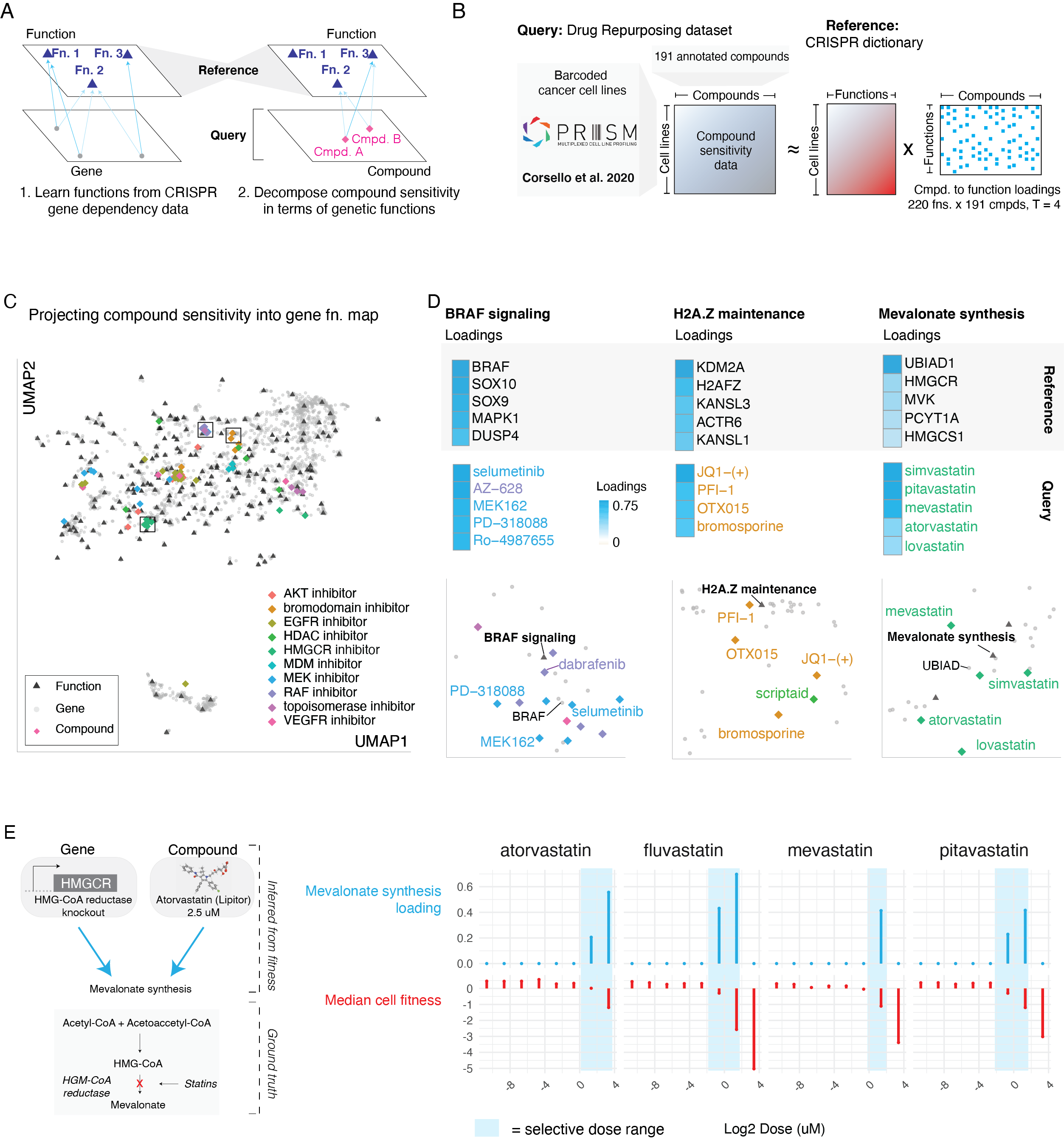}
\caption{\textit{Projecting compound perturbations into a reference space of gene functions.} A.	Conceptual overview of reference-query projection for fitness data. If Webster learned true biological functions from gene perturbation data, those functions should generalize to unseen perturbations measured over the same cell lines, such as compound sensitivity measurements.
B.	Overview of query dataset. Compounds from the Drug Repurposing Hub were screened at a uniform dose (2.5 uM) over a set of barcoded cell lines to generate compound sensitivity profiles (Corsello et al., 2020). We modeled 191 high-variance compound treatments by approximating each compound sensitivity profile as a mixture of up to four gene functions learned from CRISPR data using Webster. Compound data was not used during the training of the gene function dictionary.
C.	Joint UMAP embedding of genes, functions, and compound sensitivity profiles. Compounds embedded near gene functions reflecting their mechanism of action.
D.	Focus on three gene functions: BRAF signaling, H2A.Z maintenance, and Mevalonate synthesis. Top: The five genes most strongly loaded onto each function are shown next to a heatmap of their loading scores. Middle: The compounds most strongly loaded onto each gene function are shown next to a heatmap of their loading scores. Bottom: Insets of the embedding shown in C centered on each of the three gene functions. 
E.	Query compound projection onto reference gene functions varies by dose. In a secondary screen, various HMGCR inhibitors were screened at an 8-point dose curve ranging from 10 uM to 8 nM. Compound sensitivity profiles at each dose point were modeled independently in terms of Webster’s gene functions. The resulting loadings on the Mevalonate synthesis function are plotted against the dose. 
}
\label{f4}
\end{figure}

\subsection{Learned functions reflected a cellular hierarchy}
Next, we explored general properties of Webster’s learned latent space. For example, gene products are spatially regulated within a cellular hierarchy (Figure $\ref{f3}$D), but because Webster is trained on fitness data alone, it has no prior knowledge of this hierarchy. We leveraged recently published proximity labeling measurements (Go et al., 2021) represented as probability distributions for each gene product over twenty subcellular locations. By multiplying gene location probabilities with our gene-to-function loadings inferred from cancer fitness screens, we computed scores indicating the level of physical compartmentalization for each fitness function (Figure $\ref{sf4}$A-B). We visualized these inferences by jointly embedding 2,921 genes and 220 functions according to their fitness effects using UMAP and coloring functions according to their localization (Figure 3E). Functions enriched for specific compartments occupied nearby regions of the embedding space (Figure $\ref{sf4}$C), with pleiotropic genes embedded as data points nearby their underlying functions (Figure $\ref{f3}$F). This echoes the hierarchical relationships observed in classic yeast fitness screens (Costanzo et al., 2016; Kramer et al., 2014).

\subsection{Projecting compounds into a latent space defined by gene functions}
Finally, we assessed whether our dictionary optimized on gene perturbation data captured generalized patterns present in unseen perturbations (Figure $\ref{f4}$A). The PRISM Drug Repurposing dataset contains compound sensitivity measurements across hundreds of cancer cell lines (Corsello et al., 2020). We isolated $\sim$200 compounds with diverse and well-annotated mechanisms of action from the PRISM dataset and modeled each compound sensitivity profile as a mixture of four elements of the dictionary trained exclusively on gene perturbation data (Figure $\ref{f4}$B, Table S3). Approximation quality of compound sensitivity using gene functions varied by compound class (Figure $\ref{sf5}$A), with clinical anticancer compounds such as MDM and EGFR inhibitors exhibiting the best approximations, and broadly cytotoxic compounds such as Aurora Kinase inhibitors exhibiting the least robust approximations (Figure $\ref{sf5}$B). This suggests that Webster learned representations of on-target pathways for certain compound classes from gene perturbation data alone, in concordance with previous results on pairwise gene-drug correlations (Gonçalves et al., 2020). Examples included RAF and MEK inhibitors projecting onto the BRAF Signaling function learned from gene perturbations, bromodomain inhibitors projecting onto the H2A.Z maintenance function, and HMGCR inhibitors projecting onto the Mevalonate Synthesis function (Figure $\ref{f4}$C-D).  Some profiles were modeled as interpretable mixtures of multiple independent biological pathways, reflecting polypharmacology (ex: ATK inhibition was modeled as mixtures of RICTOR/AKT Signaling, PIK3CA Signaling and PTEN Signaling functions) (Figure $\ref{sf5}$D). We also found that projection onto on-target pathways was sensitive to compound dose, with high doses that caused broad cytotoxicity or low doses with no fitness effect failing to project at all (Figure $\ref{f4}$E, Figure $\ref{sf5}$E).

\section{Discussion}
Wagner and Zhang nominated pleiotropic inference as an outstanding problem in functional genomics a decade ago (Wagner and Zhang, 2011), envisioning two major challenges: deriving “biologically non-arbitrary” latent phenotypes from high dimensional data, and accounting for the sparse nature of genotype-phenotype relationships (Wang et al., 2010). By framing pleiotropy as an instance of the sparse representation problem (Elad, 2010) solved by graph-regularized dictionary learning, we recovered interpretable latent dictionary elements that sparsely combine to model gene effects, thereby satisfying the two challenges outlined above. Webster meets the  analytical challenge posed by growing CRISPR-Cas9 fitness screens by enabling unsupervised learning of biological functions using only numerical fitness data as input. It is generally applicable to fitness screen collections of various designs, and may be especially useful in organisms in which fitness screens are experimentally tractable but gene functions are poorly annotated, such as bacteria (Price et al., 2018). Future work could address limitations of our current method, for instance by regularizing on hierarchical graph structures rather than simple neighbor graphs, or by learning the degree of pleiotropy as a per-gene parameter rather than setting it globally.

Representing gene effects as vectors distributed over a space of latent functions runs counter to traditional symbolic representations (ex: Gene A performs Function B if Condition C is satisfied) (Fraser and Marcotte, 2004). However, in natural language processing, word symbols are represented as vectors distributed over a space of latent semantics (Mikolov et al., 2013; Pennington et al., 2014). Dictionary learning applied to word vectors (word2Vec, GloVe) resolves polysemic words as sparse linear combinations of latent meanings, such as:
$apple \approx  0.16 * fruit + 0.22 * mobile\&IT + \ldots$ (Zhang et al., 2019).

The fact that polysemic words and pleiotropic genes are modeled by dictionary learning suggests that word co-occurrence and gene co-fitness share statistical regularities. Perhaps the distributional hypothesis of semantics (“you shall know a word by the company it keeps”) (Boleda, 2020) also applies to gene function (“you shall know a gene by shared causal effects with other genes”). If so, it may be advantageous to transition from genotype-phenotype “maps” to “geometries”, where statistically independent phenotypes form an empirically derived latent space in which gene effects are vectors.
\newpage

\section{References}
Ameziane, N., May, P., Haitjema, A., van de Vrugt, H.J., van Rossum-Fikkert, S.E., Ristic, D., Williams, G.J., Balk, J., Rockx, D., Li, H., et al. (2015). A novel Fanconi anaemia subtype associated with a dominant-negative mutation in RAD51. Nat. Commun. 6, 8829.

Amici, D.R., Jackson, J.M., Truica, M.I., Smith, R.S., Abdulkadir, S.A., and Mendillo, M.L. (2021). FIREWORKS: a bottom-up approach to integrative coessentiality network analysis. Life Sci Alliance 4.

Aregger, M., Lawson, K.A., Billmann, M., Costanzo, M., Tong, A.H.Y., Chan, K., Rahman, M., Brown, K.R., Ross, C., Usaj, M., et al. (2020). Systematic mapping of genetic interactions for de novo fatty acid synthesis identifies C12orf49 as a regulator of lipid metabolism. Nat Metab 2, 499–513.

Barghout, S.H., Aman, A., Nouri, K., Blatman, Z., Arevalo, K., Thomas, G.E., MacLean, N., Hurren, R., Ketela, T., Saini, M., et al. (2021). A genome-wide CRISPR/Cas9 screen in acute myeloid leukemia cells identifies regulators of TAK-243 sensitivity. JCI Insight.

Barish, S., Barakat, T.S., Michel, B.C., Mashtalir, N., Phillips, J.B., Valencia, A.M., Ugur, B., Wegner, J., Scott, T.M., Bostwick, B., et al. (2020). BICRA, a SWI/SNF Complex Member, Is Associated with BAF-Disorder Related Phenotypes in Humans and Model Organisms. Am. J. Hum. Genet. 107, 1096–1112.

Bayraktar, E.C., La, K., Karpman, K., Unlu, G., Ozerdem, C., Ritter, D.J., Alwaseem, H., Molina, H., Hoffmann, H.-H., Millner, A., et al. (2020). Metabolic coessentiality mapping identifies C12orf49 as a regulator of SREBP processing and cholesterol metabolism. Nat Metab 2, 487–498.

Behan, F.M., Iorio, F., Picco, G., Goncalves, E., Beaver, C.M., Migliardi, G., Santos, R., Rao, Y., Sassi, F., Pinnelli, M., et al. (2019). Prioritization of cancer therapeutic targets using CRISPR-Cas9 screens. Nature 568, 511–516.

Boeing, S., Williamson, L., Encheva, V., Gori, I., Saunders, R.E., Instrell, R., Aygun, O., Rodriguez-Martinez, M., Weems, J.C., Kelly, G.P., et al. (2016). Multiomic Analysis of the UV-Induced DNA Damage Response. Cell Rep. 15, 1597–1610.

Boleda, G. (2020). Distributional Semantics and Linguistic Theory. Annu. Rev. Linguist. 6, 213–234.

Boyle, E.A., Pritchard, J.K., and Greenleaf, W.J. (2018). High-resolution mapping of cancer cell networks using co-functional interactions. Mol. Syst. Biol. 14, e8594.

Cleary, B., Cong, L., Cheung, A., Lander, E.S., and Regev, A. (2017). Efficient Generation of Transcriptomic Profiles by Random Composite Measurements. Cell 171, 1424–1436 e18.

Colic, M., Wang, G., Zimmermann, M., Mascall, K., McLaughlin, M., Bertolet, L., Lenoir, W.F., Moffat, J., Angers, S., Durocher, D., et al. (2019). Identifying chemogenetic interactions from CRISPR screens with drugZ. Genome Med. 11, 52.

Corsello, S.M., Nagari, R.T., Spangler, R.D., Rossen, J., Kocak, M., Bryan, J.G., Humeidi, R., Peck, D., Wu, X., Tang, A.A., et al. (2020). Discovering the anti-cancer potential of non-oncology drugs by systematic viability profiling. Nat Cancer 1, 235–248.

Costanzo, M., Baryshnikova, A., Bellay, J., Kim, Y., Spear, E.D., Sevier, C.S., Ding, H., Koh, J.L.Y., Toufighi, K., Mostafavi, S., et al. (2010). The genetic landscape of a cell. Science 327, 425–431.

Costanzo, M., VanderSluis, B., Koch, E.N., Baryshnikova, A., Pons, C., Tan, G., Wang, W., Usaj, M., Hanchard, J., Lee, S.D., et al. (2016). A global genetic interaction network maps a wiring diagram of cellular function. Science 353.

Costanzo, M., Kuzmin, E., van Leeuwen, J., Mair, B., Moffat, J., Boone, C., and Andrews, B. (2019). Global Genetic Networks and the Genotype-to-Phenotype Relationship. Cell 177, 85–100.

Costello, J.L., Castro, I.G., Hacker, C., Schrader, T.A., Metz, J., Zeuschner, D., Azadi, A.S., Godinho, L.F., Costina, V., Findeisen, P., et al. (2017). ACBD5 and VAPB mediate membrane associations between peroxisomes and the ER. J. Cell Biol. 216, 331–342.

Dempster, J.M., Rossen, J., Kazachkova, M., Pan, J., Kugener, G., Root, D.E., and Tsherniak, A. (2019). Extracting Biological Insights from the Project Achilles Genome-Scale CRISPR Screens in Cancer Cell Lines.

Drew, K., Wallingford, J.B., and Marcotte, E.M. (2020). hu.MAP 2.0: Integration of over 15,000 proteomic experiments builds a global compendium of human multiprotein assemblies.

Dudley, A.M., Janse, D.M., Tanay, A., Shamir, R., and Church, G.M. (2005). A global view of pleiotropy and phenotypically derived gene function in yeast. Mol. Syst. Biol. 1, 2005.0001.

Elad, M. (2010). Sparse and Redundant Representations: From Theory to Applications in Signal and Image Processing (Springer Science \& Business Media).

Fraser, A.G., and Marcotte, E.M. (2004). A probabilistic view of gene function. Nat. Genet. 36, 559–564.

Go, C.D., Knight, J.D.R., Rajasekharan, A., Rathod, B., Hesketh, G.G., Abe, K.T., Youn, J.-Y., Samavarchi-Tehrani, P., Zhang, H., Zhu, L.Y., et al. (2021). A proximity-dependent biotinylation map of a human cell. Nature.

Gonçalves, E., Segura-Cabrera, A., Pacini, C., Picco, G., Behan, F.M., Jaaks, P., Coker, E.A., van der Meer, D., Barthorpe, A., Lightfoot, H., et al. (2020). Drug mechanism-of-action discovery through the integration of pharmacological and CRISPR screens. Mol. Syst. Biol. 16, e9405.

Gratten, J., and Visscher, P.M. (2016). Genetic pleiotropy in complex traits and diseases: implications for genomic medicine. Genome Med. 8, 78.

Hart, T., Chandrashekhar, M., Aregger, M., Steinhart, Z., Brown, K.R., MacLeod, G., Mis, M., Zimmermann, M., Fradet-Turcotte, A., Sun, S., et al. (2015). High-Resolution CRISPR Screens Reveal Fitness Genes and Genotype-Specific Cancer Liabilities. Cell 163, 1515–1526.

Henkel, L., Rauscher, B., and Boutros, M. (2019). Context-dependent genetic interactions in cancer. Curr. Opin. Genet. Dev. 54, 73–82.

Hesketh, G.G., Papazotos, F., and Pawling, J. (2020). The GATOR–Rag GTPase pathway inhibits mTORC1 activation by lysosome-derived amino acids.

Hou, L., Wang, Y., Liu, Y., Zhang, N., Shamovsky, I., Nudler, E., Tian, B., and Dynlacht, B.D. (2019). Paf1C regulates RNA polymerase II progression by modulating elongation rate. Proc. Natl. Acad. Sci. U. S. A. 116, 14583–14592.

Hua, R., Cheng, D., Coyaud, É., Freeman, S., Di Pietro, E., Wang, Y., Vissa, A., Yip, C.M., Fairn, G.D., Braverman, N., et al. (2017). VAPs and ACBD5 tether peroxisomes to the ER for peroxisome maintenance and lipid homeostasis. J. Cell Biol. 216, 367–377.

Hustedt, N., Álvarez-Quilón, A., McEwan, A., Yuan, J.Y., Cho, T., Koob, L., Hart, T., and Durocher, D. (2019). A consensus set of genetic vulnerabilities to ATR inhibition. bioRxiv 574533.

Kairov, U., Cantini, L., Greco, A., Molkenov, A., Czerwinska, U., Barillot, E., and Zinovyev, A. (2017). Determining the optimal number of independent components for reproducible transcriptomic data analysis. BMC Genomics 18, 1–13.

Keeling, D.M., Garza, P., Nartey, C.M., and Carvunis, A.-R. (2019). The meanings of “function” in biology and the problematic case of de novo gene emergence. Elife 8.

Kim, H., and Park, H. (2007). Sparse non-negative matrix factorizations via alternating non-negativity-constrained least squares for microarray data analysis. Bioinformatics 23, 1495–1502.

Kim, E., Dede, M., Lenoir, W.F., Wang, G., Srinivasan, S., Colic, M., and Hart, T. (2019). A network of human functional gene interactions from knockout fitness screens in cancer cells. Life Sci Alliance 2.

Kim, E., Gheorge, V., and Hart, T. (2021). Dynamic rewiring of biological activity across genotype and lineage revealed by context-dependent functional interactions.

Kinsler, G., Geiler-Samerotte, K., and Petrov, D.A. (2020). Fitness variation across subtle environmental perturbations reveals local modularity and global pleiotropy of adaptation. Elife 9.

Koch, E.N., Costanzo, M., Deshpande, R., Andrews, B., Boone, C., and Myers, C.L. (2017). Systematic identification of pleiotropic genes from genetic interactions.

Li, H., Ning, S., Ghandi, M., Kryukov, G.V., Gopal, S., Deik, A., Souza, A., Pierce, K., Keskula, P., Hernandez, D., et al. (2019). The landscape of cancer cell line metabolism. Nat. Med. 25, 850–860.

Lightfoot, H.L., Hagen, T., Cléry, A., Allain, F.H.-T., and Hall, J. (2018). Control of the polyamine biosynthesis pathway by G2-quadruplexes. Elife 7.

Loregger, A., Raaben, M., Nieuwenhuis, J., Tan, J.M.E., Jae, L.T., van den Hengel, L.G., Hendrix, S., van den Berg, M., Scheij, S., Song, J.Y., et al. (2020). Haploid genetic screens identify SPRING/C12ORF49 as a determinant of SREBP signaling and cholesterol metabolism. Nat. Commun. 11, 1128.

Mairal, J., Bach, F., and Ponce, J. (2014). Sparse Modeling for Image and Vision Processing.

Malovannaya, A., Li, Y., Bulynko, Y., Jung, S.Y., Wang, Y., Lanz, R.B., O’Malley, B.W., and Qin, J. (2010). Streamlined analysis schema for high-throughput identification of endogenous protein complexes. Proc. Natl. Acad. Sci. U. S. A. 107, 2431–2436.

Mashtalir, N., D’Avino, A.R., Michel, B.C., Luo, J., Pan, J., Otto, J.E., Zullow, H.J., McKenzie, Z.M., Kubiak, R.L., St Pierre, R., et al. (2018). Modular Organization and Assembly of SWI/SNF Family Chromatin Remodeling Complexes. Cell 175, 1272–1288 e20.

McDonald, E.R., 3rd, de Weck, A., Schlabach, M.R., Billy, E., Mavrakis, K.J., Hoffman, G.R., Belur, D., Castelletti, D., Frias, E., Gampa, K., et al. (2017). Project DRIVE: A Compendium of Cancer Dependencies and Synthetic Lethal Relationships Uncovered by Large-Scale, Deep RNAi Screening. Cell 170, 577–592.e10.

McInnes, L., Healy, J., and Melville, J. (2018). UMAP: Uniform Manifold Approximation and Projection for Dimension Reduction.

Meyers, R.M., Bryan, J.G., McFarland, J.M., Weir, B.A., Sizemore, A.E., Xu, H., Dharia, N.V., Montgomery, P.G., Cowley, G.S., Pantel, S., et al. (2017). Computational correction of copy-number effect improves specificity of CRISPR-Cas9 essentiality screens in cancer cells. Nat. Genet.

Michel, B.C., D’Avino, A.R., Cassel, S.H., Mashtalir, N., McKenzie, Z.M., McBride, M.J., Valencia, A.M., Zhou, Q., Bocker, M., Soares, L.M.M., et al. (2018). A non-canonical SWI/SNF complex is a synthetic lethal target in cancers driven by BAF complex perturbation. Nat. Cell Biol. 20, 1410–1420.

Mikolov, T., Yih, W.-T., and Zweig, G. (2013). Linguistic Regularities in Continuous Space Word Representations. In Proceedings of the 2013 Conference of the North American Chapter of the Association for Computational Linguistics: Human Language Technologies, (Atlanta, Georgia: Association for Computational Linguistics), pp. 746–751.

Olivieri, M., Cho, T., Álvarez-Quilón, A., Li, K., Schellenberg, M.J., Zimmermann, M., Hustedt, N., Rossi, S.E., Adam, S., Melo, H., et al. (2020). A Genetic Map of the Response to DNA Damage in Human Cells. Cell 182, 481–496.e21.

Pan, J., Meyers, R.M., Michel, B.C., Mashtalir, N., Sizemore, A.E., Wells, J.N., Cassel, S.H., Vazquez, F., Weir, B.A., Hahn, W.C., et al. (2018). Interrogation of Mammalian Protein Complex Structure, Function, and Membership Using Genome-Scale Fitness Screens. Cell Syst 6, 555–568 e7.

Pati, Y.C., Rezaiifar, R., and Krishnaprasad, P.S. (1993). Orthogonal matching pursuit: recursive function approximation with applications to wavelet decomposition. In Proceedings of 27th Asilomar Conference on Signals, Systems and Computers, pp. 40–44 vol.1.

Pennington, J., Socher, R., and Manning, C.D. (2014). Glove: Global vectors for word representation. In Proceedings of the 2014 Conference on Empirical Methods in Natural Language Processing (EMNLP), pp. 1532–1543.

Price, M.N., Wetmore, K.M., Waters, R.J., Callaghan, M., Ray, J., Liu, H., Kuehl, J.V., Melnyk, R.A., Lamson, J.S., Suh, Y., et al. (2018). Mutant phenotypes for thousands of bacterial genes of unknown function. Nature 557, 503–509.

Rahman, M., Billmann, M., Costanzo, M., Aregger, M., Tong, A.H.Y., Chan, K., Ward, H.N., Brown, K.R., Andrews, B.J., Boone, C., et al. (2021). A method for benchmarking genetic screens reveals a predominant mitochondrial bias. Mol. Syst. Biol. 17, e10013.

Rancati, G., Moffat, J., Typas, A., and Pavelka, N. (2018). Emerging and evolving concepts in gene essentiality. Nat. Rev. Genet. 19, 34–49.

Raudvere, U., Kolberg, L., Kuzmin, I., Arak, T., Adler, P., Peterson, H., and Vilo, J. (2019). g:Profiler: a web server for functional enrichment analysis and conversions of gene lists (2019 update). Nucleic Acids Res. 47, W191–W198.

Rauscher, B., Heigwer, F., Henkel, L., Hielscher, T., Voloshanenko, O., and Boutros, M. (2018). Toward an integrated map of genetic interactions in cancer cells. Mol. Syst. Biol. 14, e7656.

Rubinstein, R., Bruckstein, A.M., and Elad, M. (2010). Dictionaries for Sparse Representation Modeling. Proc. IEEE 98, 1045–1057.

Sanchez-Vega, F., Mina, M., Armenia, J., Chatila, W.K., Luna, A., La, K.C., Dimitriadoy, S., Liu, D.L., Kantheti, H.S., Saghafinia, S., et al. (2018). Oncogenic Signaling Pathways in The Cancer Genome Atlas. Cell 173, 321–337.e10.

Sanson, K.R., Hanna, R.E., Hegde, M., Donovan, K.F., Strand, C., Sullender, M.E., Vaimberg, E.W., Goodale, A., Root, D.E., Piccioni, F., et al. (2018). Optimized libraries for CRISPR-Cas9 genetic screens with multiple modalities. Nat. Commun. 9, 5416.

Sekelsky, J.J., Burtis, K.C., and Hawley, R.S. (1998). Damage control: the pleiotropy of DNA repair genes in Drosophila melanogaster. Genetics 148, 1587–1598.

Solovieff, N., Cotsapas, C., Lee, P.H., Purcell, S.M., and Smoller, J.W. (2013). Pleiotropy in complex traits: challenges and strategies. Nat. Rev. Genet. 14, 483–495.

Stein-O’Brien, G.L., Arora, R., Culhane, A.C., Favorov, A.V., Garmire, L.X., Greene, C.S., Goff, L.A., Li, Y., Ngom, A., Ochs, M.F., et al. (2018). Enter the Matrix: Factorization Uncovers Knowledge from Omics. Trends Genet. 34, 790–805.

Szklarczyk, D., Franceschini, A., Wyder, S., Forslund, K., Heller, D., Huerta-Cepas, J., Simonovic, M., Roth, A., Santos, A., Tsafou, K.P., et al. (2014). STRING v10: protein–protein interaction networks, integrated over the tree of life. Nucleic Acids Res. 43, D447–D452.

Tsai, K.-L., Tomomori-Sato, C., Sato, S., Conaway, R.C., Conaway, J.W., and Asturias, F.J. (2014). Subunit architecture and functional modular rearrangements of the transcriptional mediator complex. Cell 157, 1430–1444.

Tsherniak, A., Vazquez, F., Montgomery, P.G., Weir, B.A., Kryukov, G., Cowley, G.S., Gill, S., Harrington, W.F., Pantel, S., Krill-Burger, J.M., et al. (2017). Defining a Cancer Dependency Map. Cell 170, 564–1916796928.

Tyler, A.L., Crawford, D.C., and Pendergrass, S.A. (2016). The detection and characterization of pleiotropy: discovery, progress, and promise. Brief. Bioinform. 17, 13–22.

Wagner, G.P., and Zhang, J. (2011). The pleiotropic structure of the genotype-phenotype map: the evolvability of complex organisms. Nat. Rev. Genet. 12, 204–213.

Wainberg, M., Kamber, R.A., Balsubramani, A., Meyers, R.M., Sinnott-Armstrong, N., Hornburg, D., Jiang, L., Chan, J., Jian, R., Gu, M., et al. (2021). A genome-wide atlas of co-essential modules assigns function to uncharacterized genes. Nat. Genet. 1–12.

Wang, T., Yu, H., Hughes, N.W., Liu, B., Kendirli, A., Klein, K., Chen, W.W., Lander, E.S., and Sabatini, D.M. (2017). Gene Essentiality Profiling Reveals Gene Networks and Synthetic Lethal Interactions with Oncogenic Ras. Cell 168, 890–903.e15.

Wang, Z., Liao, B.-Y., and Zhang, J. (2010). Genomic patterns of pleiotropy and the evolution of complexity. Proc. Natl. Acad. Sci. U. S. A. 107, 18034–18039.

Watanabe, K., Stringer, S., Frei, O., Umićević Mirkov, M., de Leeuw, C., Polderman, T.J.C., van der Sluis, S., Andreassen, O.A., Neale, B.M., and Posthuma, D. (2019). A global overview of pleiotropy and genetic architecture in complex traits. Nat. Genet. 51, 1339–1348.

Xiao, J., Xiong, Y., Yang, L.-T., Wang, J.-Q., Zhou, Z.-M., Dong, L.-W., Shi, X.-J., Zhao, X., Luo, J., and Song, B.-L. (2020). POST1/C12ORF49 regulates the SREBP pathway by promoting site-1 protease maturation. Protein Cell.

Yankelevsky, Y., and Elad, M. (2016). Dual Graph Regularized Dictionary Learning. IEEE Transactions on Signal and Information Processing over Networks 2, 611–624.

Yankelevsky, Y., and Elad, M. (2020). Theoretical guarantees for graph sparse coding. Appl. Comput. Harmon. Anal. 49, 698–725.

Zhang, J., Chen, Y., Cheung, B., and Olshausen, B.A. (2019). Word Embedding Visualization Via Dictionary Learning.

\section{Acknowledgements}

 Work was funded by U01 CA176058 (W.C.H.). We thank members of the Hahn Lab and the Cancer Data Science team, as well as Caleb Laureau, Peter Winter, Kevin Zhan, Luc de Waal, Yael Yankelevsky and Niklas Rindorff for helpful feedback and discussions. Thank you to Ben Cooley, Andrew Tang, Nena Berg and Kat Huang of the Pattern Team @ Broad (https://pattern.broadinstitute.org/) for developing the minisite. Thank you to Mariya Kazachkova for sharing her arm correction script. Figures created with assets from BioRender.com.

\newpage

\appendix
\section{Appendix}
\label{appendix}

\setcounter{figure}{0} \renewcommand{\thefigure}{S\arabic{figure}}
\section{Data and code availability}

We provide two code repositories and two data repositories to support this manuscript. The main repository contains R code for reproducing figures and analyses presented in the paper, and can be found at \url{https://github.com/joshbiology/gene_fn}. We created a Figshare archive that is the starting point for these analyses (\url{https://www.doi.org/10.6084/m9.figshare.14960006}). We provide a second repository of MATLAB code implementing the factorization methods that are the basis of Webster. This can be found at \url{https://github.com/joshbiology/graph_dictionary_learning}. Finally, we provide a data repository containing Webster’s preprocessed numerical input data and subsequent analysis output, a subset of which form the Supplemental Tables cited throughout this paper (\url{https://www.doi.org/10.6084/m9.figshare.14963561}). 

\section{Method details}
\subsection{Overview of the basic sparse approximation problem}
The field of sparse approximation encompasses a diverse range of applications and implementation strategies. One specific formulation of the sparse approximation problem is centered around the following task (from (Elad, 2010), “Chapter 9.2 The Sparse-Land Model”, see also (Yankelevsky and Elad, 2016), “I. Introduction” and (Mairal et al., 2014), “Chapter 1.4 Dictionary Learning”):

Suppose a set of $m$ training signals $Y = [y_1, y_2, \ldots, y_m]$ in $\mathbb{R}^n$. From this set of training signals alone, we seek to approximate them by recovering:

(1)	A dictionary matrix $D$ in $\mathbb{R}^{n \times k}$;

(2)	Sparse representations of each training signal in terms of dictionary elements. These representations form a matrix $X = [x_1, x_2, \ldots, x_m]$ with each $x_i$ being a $t$-sparse vector in $\mathbb{R}^k$.

The approximation of each training signal is $y_i \approx D * x_i$. The term on the left is the measured signal, which is a vector in $\mathbb{R}^n$. The term on the right is a matrix operation resulting in a vector in $\mathbb{R}^n$, composed from the weighted sum of $t$ columns of $D$ (each of which are vectors in $\mathbb{R}^n$), with the weights being the $t$ non-zero coefficients held in $x_i$.

We can also express this in matrix form: $Y \approx D * X$.

To solve the sparse approximation problem, we find an optimal solution to both $D$ and $X$ that minimizes the approximation error, given the sparsity constraints imposed by the parameter $t$, and the number of dictionary elements dictated by $k$. Formally, the objective is

\makebox[\linewidth]{$\underset{D,X}{\text{arg\,min}} ||Y-DX||^2_F$ s.t. $||x_i||_0 \leq t$  $\forall i$}
 
\subsection{Sparse approximation of pleiotropic gene functions from fitness data}
We adapt the framework of sparse approximation to model fitness data. Fitness data consists of measured changes in cell growth rate upon biological perturbations applied across cell contexts. Here, we focus on gene perturbation, which can be induced in a variety of ways, including with programmable CRISPR-Cas9 nucleases. For additional descriptions of fitness experiments and their design, see (Rancati et al., 2018) (focused on human cells) and (Costanzo et al., 2019) (focused primarily on yeast).

We are specifically interested in modeling the genetic architecture underlying fitness measurements. In particular, we seek to model pleiotropy, in which gene products have multiple functions depending on the cell context. Solving pleiotropy within fitness data reduces to (1) learning a set of fitness effects of perturbed biological functions, and (2) modeling gene effects as a combination of these functional effects. 

There is a clear connection between these two tasks and the basic sparse approximation problem described above, although augmentations are required to encourage latent dictionary elements to model biological functions rather than numerically valid -- but biologically meaningless -- patterns in the data. To achieve this, we exploit several properties of fitness data. First, genes with correlated fitness effects tend to have similar biological functions, a concept referred to as co-essentiality (Amici et al., 2021; Bayraktar et al., 2020; Boyle et al., 2018; Kim et al., 2019; Pan et al., 2018; Wainberg et al., 2021; Wang et al., 2017). Second, similar cell contexts will share similar rate limiting functions for cell growth. This concept has been most extensively explored in the cancer field, in which related cell lines have been shown to harbor similar “selective dependencies” (McDonald et al., 2017; Tsherniak et al., 2017). Constraining learned representations to capture coherent gene modules essential in specific cell contexts is consistent with past use of bi-clustering to model pleiotropy (Dudley et al., 2005; Koch et al., 2017).

In summary, to recover latent representations in our model that capture biological functions, we  augmented the basic sparse approximation objective (error minimization) with two additional objectives: learned representations should preserve the local structure between gene effects as well as preserve the local structure between cell contexts. To operationalize this, we adopted the framework of dual-graph regularized dictionary learning.

\subsection{Webster: dual graph regularized dictionary learning applied to preprocessed fitness data} 
Our overall approach (‘Webster’) for modeling pleiotropic gene functions in fitness data consists of two steps. First, we preprocess the fitness data. This is done by centering and scaling individual screens, then centering the fitness effects of each gene perturbation (‘gene effects’), and finally filtering out low variance gene effects. Batch correction and other quality control steps are applied at this stage as well, described in detail in later sections. Second, from this matrix of preprocessed gene effects, Webster uses dual graph regularized dictionary learning (DGRDL) (Yankelevsky and Elad, 2016, 2020) to sparsely approximate the gene effect matrix in terms of latent biological functions. 

Mathematically, this can be described as follows. Let the set of gene effects $Y = [y_1, y_2, \ldots, y_m]$ in $\mathbb{R}^n$ consist of fitness effects upon individually perturbing m genes across n cell contexts. Pairwise similarities between the m gene effects define a matrix $W$ in $\mathbb{R}^{m \times m}$, while the pairwise similarities between the $n$ cell contexts define a matrix $V$ in $\mathbb{R}^{n \times n}$. 

From $Y$, $W$ and $V$, we recover a dictionary matrix $D$ in $\mathbb{R}^{n \times k}$ and a sparse representation matrix $X$ in $\mathbb{R}^{k \times m}$ (with each column of X containing only $t$ non-zero entries). $D$ and $X$ are optimized to satisfy the following objectives:

(1)	Minimize the total approximation error $|| Y - D * X ||_F$; 

(2)	Minimize the sum of squared differences between the columns of $X$, weighted by the similarities of the corresponding columns of $Y$, expressed as $\frac{1}{2} \sum_{  1 \leq i, j \leq m} W_{i,j} * ( X[ , i]  - X[ , j] )^2$;

(3)	Minimize the sum of squared differences between the rows of $D$, weighted by the similarities of the corresponding rows of Y, expressed as $\frac{1}{2} \sum_{  1 \leq i, j \leq n} V_{i,j} * ( D[i, ]  - D[ j, ] )^2$.

The intuition behind objective (2) is that the more similar two gene effects are in the original data, the more similar their sparse representations should be; i.e. they should be approximated using similar biological functions. The intuition behind objective (3) is that the more similar two cell contexts are in their growth requirements, the more similar their representations in the dictionary matrix should be, i.e they should depend on the same functions for growth. Formally, objectives (2) and (3) can be compactly expressed as a quadratic form of a graph Laplacian derived from the gene effect similarity graph and the cell context similarity graph, respectively.

The final DGRDL objective function is:

\makebox[\linewidth]{$\underset{D,X}{\text{arg\,min}} ||Y-DX||^2_F + \alpha Tr(D^T L D) + \beta Tr(X L_c X^T)$ s.t. $||x_i||_0 \leq t$  $\forall i$}

where $L$ in $\mathbb{R}^{m \times m}$ is the Laplacian matrix derived from the cell context similarity graph, $L_c$ in $\mathbb{R}^{n \times n}$ is the Laplacian of the gene effect similarity graph, and $\alpha$ and $\beta$ are importance weights for each term in the final objective. In practice, we choose $k < m$, as we assume that gene effects can be described by a small set of latent elements. We also tend to choose $k < n$, resulting in an undercomplete dictionary basis. Finally, we choose $t$ to be small ($t << k$) but greater than $1$.

This is a high-level summary of DGRDL specifically through the lens of fitness data. For a fuller exploration of DGRDL applied to other data modalities, as well as theoretical guarantees of DGRDL for signal recovery, see the original papers: (Yankelevsky and Elad, 2016, 2020).

\subsection{Implementation details}
In order to implement DGRDL, we obtained MATLAB code for k-SVD, orthogonal matching pursuit, and DGRDL from the respective lab websites (\url{https://elad.cs.technion.ac.il/software/} and \url{http://www.cs.technion.ac.il/~ronrubin/software.html}), which we detailed in the documentation for our code repository (\url{https://github.com/joshbiology/graph_dictionary_learning}).

Our Webster approach customizes the base DGRDL method in several ways. First, we use nearest-neighbor graphs in our graph regularization terms, with the goal of preserving local structure present in the data. Empirically, nearest neighbor graphs capture the overall topological relationships present in gene effect similarity networks (Amici et al., 2021; Pan et al., 2018), and provide the highest enrichment for previously annotated gene-gene relationships (Boyle et al., 2018; Kim et al., 2019). As a side note, nearest neighbor graphs are also the input to popular manifold learning algorithms t-SNE and UMAP, which tend to sufficiently capture local structure in biological data. We use nearest neighbor gene similarity and cell context similarity graphs as the basis for the graph regularization objectives described above, using cosine similarity and five nearest neighbors for both graphs (unless otherwise specified below). 

Our second customization involves the dictionary learning initialization step. As DGRDL is an iterative optimization algorithm, it requires a pre-initialized dictionary in order to perform its first iteration. In the absence of a pre-initialized dictionary, DGRDL chooses $k$ random elements among the input training signals to serve as the initial dictionary. We found empirically that this random selection introduced variability in the optimization outcomes. As a result, we pass a pre-initialized dictionary to DGRDL in the form of $k$ representative training signals chosen via $k$-medoids from among the input training signals, using the MATLAB kmedoids function. As $k$-medoids is a clustering algorithm, the initial dictionary can be thought of as optimal for clustering the data into $k$ mutually exclusive groups. From this starting point, an new dictionary is learned that best approximates all m training signals in terms of $t$ dictionary elements while preserving local structure of the data. As $t >1$, this process can be thought of as relaxing the one-hot clustering assumptions present in the initial dictionary. 

Finally, we multiply the sparse representation matrix $X$ with a scalar correction factor $1/\sqrt n$ to convert coefficients into units of standard deviation from their original unit of Euclidean distance. As these coefficients are analogous to the loading coefficients used in PCA, we refer to this matrix of coefficients as the loadings matrix in the manuscript.

\subsection{Fitness data origins and preprocessing}
The genotoxic screening collection was downloaded from the publisher's website (Olivieri et al., 2020). The dataset consisted of 17,382 measured gene effects over 31 fitness screens performed in the presence of low doses of genotoxins. Each of the 31 screens was already processed through normZ which centers and scales the data (Colic et al., 2019). 

To select high variance genes, we assumed that a large portion of gene effects in genome-scale screens have no phenotype in any cell contexts (also known as non-essential genes). Any variance present in these gene effects will be due to experimental noise rather than biological signal. The distribution of these across-cell-line variances will follow a chi-squared distribution, which converges to a normal distribution with large $n$. Biologically driven gene effects will exhibit greater variation across cell lines and will form positive outliers in this normal distribution. To detect such outliers, we used a quantile-quantile plot to visualize the observed distribution of gene effect variances compared to a theoretical normal distribution. We drew a cutoff that separated high variance genes from the remaining (variance $>$ 3), resulting in 304 gene effects over 31 screens that formed the input to graph regularized dictionary learning.

The 19Q4 Broad Cancer Dependency Map screening collection was downloaded from the Cancer Dependency Map FigShare archive (under the file Achilles\_gene\_effect.csv, \url{https://doi.org/10.6084/m9.figshare.11384241.v3}). The raw dataset consisted of 18,333 gene effect measurements over 689 cancer cell lines. We filtered out a group of cell lines that suffered a batch effect due to PCR contamination. We filtered out a group of X-chromosomal genes whose copy number across cell lines could not be properly controlled for. We also filtered out two HUGO gene groups, olfactory genes (\url{https://www.genenames.org/data/genegroup/#!/group/141}) and KRTAP genes (\url{https://www.genenames.org/data/genegroup/#!/group/619}), whose high sequence similarity resulted in large-scale off target guide cutting activity as previously described (Boyle et al., 2018). This intermediate dataset consisted of 17,167 gene effect measurements over 675 cell lines.

We applied several correction measures before gene effect selection. The first was a generalized correction for cutting effects specific to chromosome arms, as reported previously (Amici et al., 2021). We then centered each cell line screen and applied a batch correction for screen quality. This was done by linearly regressing out the NNMD profile over cell lines from each gene effect profile. NNMD stands for null-normalized mean difference and was previously described (Dempster et al., 2019). The NNMD score for each cell line was obtained from the “NNMD” column the sample\_info.csv file from Figshare archive (\url{https://doi.org/10.6084/m9.figshare.11384241.v3}). After screen quality correction, we scaled each cell line screen. This dataset was used as input for gene effect selection.

For gene effect selection, we used three criteria: 

(1)	Variance 

(2)	Perturbation confidence 

(3)	Maximum pairwise correlation with other gene effects

The cutoff for variance was chosen using a quantile-quantile plot, as with the genotoxic screening data above. A variance cutoff of 1 was used. The perturbation confidence score was calculated as described previously (\url{http://archive.today/2021.03.22-122633/https://cancerdatascience.org/blog/posts/gene_confidence_blog/}). In brief, an XGBoost model was trained to discriminate between known low confidence gene perturbations and high confidence gene perturbations using various features derived from the individual guide RNA effects. The recommended cutoff of 0.5 was applied. From these gene effects, we kept those that exhibited a maximum pairwise correlation with other gene effects above a certain threshold, as done in yeast fitness screen analyses (Costanzo et al., 2010, 2016). The threshold we used was 0.275. This resulted in a preprocessed matrix of 2,921 gene effect measurements over 675 cell lines that formed the input to graph regularized dictionary learning.

Sanger Institute screens processed through the CERES algorithm were downloaded from the Cancer Dependency Map FigShare archive (under the filename sample\_info.csv, \url{https://doi.org/10.6084/m9.figshare.9116732.v2}). The raw data consisted of 17,716 gene effect measurements over 318 cell lines. Arm correction, centering, NNMD regression and scaling were applied.

\subsection{Hyperparameter tuning and model selection}
For the genotoxic fitness screens, we empirically chose the primary dictionary learning hyperparameter $k$ (dictionary size) by sweeping between values of  1 and 31 while fixing $t$ = 2. We evaluated each of the model objectives and looked for diminishing returns for the approximation error and gene Laplacian model objectives (the cell context Laplacian had a linear relationship with k, so it was less informative for model selection). Diminishing returns were reached at $k$ = 10. For a comparison, we also swept across $k$ while fixing $t$ = 1 (representing hard clustering). We also repeated these experiments without graph regularization (by setting $\alpha$ and $\beta$ to 0). 

For the cancer cell fitness screens, we performed a grid search over $k$ = 25 to 675 in steps of 25, and $t$ = 1:8, resulting in 216 model instances. From this search, we selected $t$ = 4, as this was the sparsest model parameter whose objectives remained well behaved for all values of $k$. Subsequently, we performed a second sweep over $k$ = 25 to 675 in steps of 5, with $t$ fixed at 4, resulting in 131 model instances. Diminishing returns in the approximation error and gene Laplacian model objectives were observed at $k$ = 220, which we chose for the final factorization. We confirmed that model objectives for these hyperparameter choices were stable to random seed initialization.

Various other model hyperparameters were set to the default values recommended in the original paper (Yankelevsky and Elad, 2016), and were as follows: $\alpha$ = 0.2; $\beta$ = 0.6; number of iterations = 20. Finally, we set the graph regularization terms according to the settings explained above: neighbor graph degree = 5; neighbor graph metric/edgeweight = cosine similarity.

\subsection{Dictionary learning experiments: robustness, denoising and transferability}
To assess robustness, we performed DGRDL with different random seed initializations, using the same $k$-medoids initialized dictionary. Element-wide consistency between the resulting dictionaries was assessed using Pearson correlation.

To assess the denoising properties of dictionary learning on fitness data, we created four noisy versions of the cancer cell fitness screening dataset by adding Gaussian random noise. The Gaussian random noise matrices were created using the R function rnorm with standard deviations set to a variety of values (\textit{s.d.} = 0.25, 0.5, 1, 1.5). We also randomly split the 2,921 gene effects in cancer cells into 2,191 training genes and 730 test genes. 

For each of the five datasets (original data and four noise levels), we trained DGRDL on the training gene effects only (using $k$ = 220, $t$ = 4) and subsequently modeled the corresponding unseen test genes in terms of dictionary elements (with the same noise level present in the test and training genes). Each reconstructed test gene was subsequently compared to the corresponding gene effect in the raw data (which did not have additional synthetic noise added) using Pearson correlation as a metric. We repeated this entire process five times, and kept the mean Pearson correlation for each test gene at each noise level. For each noise level, the distribution of these resulting scores were plotted as a distribution. 

For the transferability experiments, we used a dictionary trained on data measured by one institute (Broad) to model the corresponding gene effects measured by another institute (Sanger), which used different experimental conditions and CRISR-Cas9 guide RNA sequences for each gene. We took a dictionary trained on the full Broad 2,191 gene effects over 675 cell lines ($k$ = 220, $t$ = 4) and subsetted it to the 161 cell lines that were screened by both institutes. For the Broad and Sanger datasets, we modeled the corresponding 2,901 overlapping gene effects across the 161 common cell lines in terms of the same Broad-trained dictionary. The resulting approximations were assessed using Pearson correlation and were plotted as a distribution for each dataset. As a comparison, a dictionary with shuffled cell line annotations was used to model the 2,901 Sanger-measured gene effects.

This concludes the descriptions of the core algorithmic steps performed with Webster in the manuscript. The remaining sections describe the annotation and visualization steps performed on Webster output.

\section{Quantification and Statistical Analysis}
\subsection{Benchmarking}
To benchmark Webster against other latent variable models, we applied PCA, ICA and $k$-medoids to identically preprocessed fitness data, with $k$ held constant. Each method, including Webster, was run 5 times with varying random seeds with the mean of each benchmark reported. For sparsity, we reported the average Hoyer index over each gene representation. ($k$-medoids always returns a perfect score of 1, since each gene is represented by a single medoid.)

\subsection{Neighbor graph embedding to visualize gene function landscapes}
For visualizing genotoxic and cancer cell fitness screen collections, we utilized the UMAP approach (McInnes et al., 2018) as implemented by the R umap package. In each case, two matrices were concatenated and used as input: the preprocessed gene effects that served as inputs to Webster, and the dictionary matrix that is outputted by Webster. For the genotoxic screens, the 304 gene effects and 10 functional effects were plotted using the following UMAP parameters: metric = “pearson”, num\_neighbors = 15. For the cancer cell fitness screens, the 2,921 gene effects and 220 functional effects were plotted using the following UMAP parameters: metric = “pearson”, num\_neighbors = 10. All other parameters were set to default. 

\subsection{Annotation of learned functions}
Dictionary elements learned by Webster from each screen collection were annotated as follows. For each of the ten dictionary elements learned from genotoxic fitness screens, we considered, in order of priority, (1) strongly loaded genes and (2) which treatments induced a strong fitness effect. For the first of the five dictionary elements , the strongly loaded genes on each mapped to one of five classical DNA damage response pathways. These relationships were corroborated by the set of treatments that induce fitness effects in these genes. For three of the dictionary elements, only a single treatment induced a strong fitness effect on its loaded genes. These were representative of the resistance / sensitivity profiles of these specific treatments, which we corroborated by matching its strongly loaded genes against similar screens from the literature. Finally, two dictionary elements exhibited strong enrichment for common essential genes and proliferation suppressor genes in their highly loaded genes, annotations for which were taken from the literature (Colic et al., 2019; Hart et al., 2015).

For each of the 220 dictionary elements learned from the cancer cell fitness screens, we again considered, in order of importance, (1) strongly loaded genes and (2) top genomic features from models trained to predict the function’s fitness profile over cell lines (see below). For annotations of the strongly loaded genes, we used a gene annotation web service, gProfiler, that takes a ranked list of genes as input and outputs a ranked list of enriched genesets (Raudvere et al., 2019). We supplemented these annotations with corresponding ones from STRING (Szklarczyk et al., 2014). We searched the literature for additional insights and recently published corroborating papers when applicable. Finally, we performed biomarker association analysis using cell line features, described below.

Using a combination of geneset enrichments, specific literature annotations for the top loaded genes and cell line features, we were able to manually annotate each dictionary element. Of the 220 dictionary elements, 199 elements were mapped to a biological function using manual curation by the authors. Of these, 195 were annotated according to geneset enrichments among top loaded genes, and 4 were named based on a clear biomarker. Nineteen of the dictionary elements were highly loaded for essential genes, which incur fitness effects at the lower detection limit of our assay and group together in our analysis for that reason. These functions were labeled as “Common Essential (Gene)”, where the Gene was chosen to be the top loaded gene, or “Common Essential (Chr\#)”, when the common essential genes shared synteny on chromosome regions (Amici et al., 2021). Finally, two dictionary elements could not be mapped to a biological process. On deeper examination, the top loaded genes for both of these functions were perturbed using Avana CRISPR-Cas9 guides that targeted non-unique genomic sequences, suggesting that these elements represented technical factors that were separated from the remainder of the data. We labeled these as “[Gene] (unclear)”. 

\subsection{Associating fitness effects with baseline genomic features of cell lines}
Baseline genomic features were associated to each function using predictive modeling. Using the function’s inferred fitness effect across cell lines as the target, we performed random forest regression using the following features from the DepMap 21Q2 release (https://doi.org/10.6084/m9.figshare.14541774.v2):

\begin{itemize}
\item RNAseq expression
\item copy number
\item boolean mutation matrices
\item Methylation
\item Proteomics
\item Lineage
\item metabolomics data
\end{itemize}

The numerical features were normalized and the categorical features were one-hot encoded and then combined into a feature matrix. These features were then used to predict the target fitness profile using 5-fold cross validation, using Pearson correlation as the metric for model performance. We selected the top 1000 features in each fold using the f-regression metric to fit the model to the target. A final model was trained on all the data and the feature importances were extracted from that model to help determine the likely features that were most important in these predictions. In certain cases, cell lines strongly dependent on a certain function harbored interpretable predictive features; those are reported in the paper when applicable, and were used to corroborate the function name chosen as described above.

\subsection{Subcellular localization analyses}
The Human Cell Map project profiled bait-prey interactions in human cells using proximity ligation. We downloaded their subcellular localization inferences for 4,424 proteins across 20 subcellular locations (Go et al., 2021). Of these proteins, 1,463 had corresponding gene effects as part of the Webster analysis cancer cell fitness screen, so their data was filtered to a dense matrix of 1,463 protein localization profiles, and the Webster loadings matrix was filtered to the same set of 1,463 genes assigned loaded across 220 inferred biological functions.

To ask whether highly loaded genes in specific functions shared subcellular localization annotations, we took the matrix product, which resulted in a new matrix of 220 functions scored across 20 localizations. Each of the 220 rows of this matrix is the sum of the individual protein-level localization distributions weighted by their loading score on that function. This was used for the basis of subcellular localization analysis of our Webster inferred functions. 

We noted during clustering of the 20 subcellular localizations (both in the original NMF matrix as well as this new matrix product) that each could be grouped into one of seven interpretable compartments. These compartment level annotations were used throughout the paper.

Finally, we report the localization specificity of individual functions. For each of the 220 functions, the entropy of its distribution over 20 localizations was calculated using the R entropy package, and the resulting 220 entropy scores were rescaled such that lowest entropy distributions were assigned a new “Specificity” score of 1, while the highest entropy distributions were assigned a score of 0.

\subsection{Compound sensitivity data}
PRISM Primary Screen compound sensitivity data was obtained from the Drug Repurposing Hub Figshare archive (primary-screen-replicate-collapsed-logfold-change.csv, \url{https://doi.org/10.6084/m9.figshare.9393293.v4}). The raw data matrix consisted of 5,274 compound sensitivity profiles measured over 578 cell lines using a 2.5 uM dose treatment. To eliminate cell line quality effects, we subtracted a trimmed mean of each cell line profile from each compound sensitivity profile. 

To prepare the data for modeling in terms of Webster’s latent functions, we selected compound classes representing known mechanisms of action, for which there were at least 5 compound sensitivity profiles in the primary screen. This matrix contained missing values, which were imputed using the R package FastImputation. Finally, we kept those cell lines that were also screened in the CRISPR-Cas9 fitness dataset. 

The final compound sensitivity profiles with known MOA’s consisted of 191 compound sensitivity profiles over 367 cell lines. Each compound was modeled in terms of the dictionary matrix learned from CRISPR-Cas9 gene perturbation, filtered to the same 367 cell lines. The modeling was performed with orthogonal matching pursuit with t = 4, such that each compound sensitivity profile was modeled as a sparse combination of four dictionary elements.

The 191 compound sensitivity profiles were added as data points to the UMAP gene function plot by reconstructing each compound sensitivity profile using full-sized dictionary elements (675 cell lines), and using the R predict function to add the imputed profile to the previously learned R umap object. 

These steps were repeated for the PRISM Secondary Screen compound sensitivity data, in which the same compounds as above were treated at multiple dose points.

\subsection{Additional factorization of genotoxic screening data}
For PCA applied to the genotoxic screening dataset, we set the number of components to 8, which was automatically chosen according to an “elbow plot” of the eigenvalues corresponding to each component in the model, using the quickelbow function in the R package bigpca. For ICA, we set the number of components to 13, according to the MSTD score from (Kairov et al., 2017), which calculates an optimal number of stable ICA components.

To compare the Webster, PCA and ICA factorizations, we computed an individual AUROC for each geneset defined by (Olivieri et al., 2020) for each inferred component, using the roc\_auc function in the R package yardstick. This computes the enrichment specificity of each geneset across components, according to the rank order of genes by their loading scores on that component.

\section{Supplemental Tables}
Table S1: Webster output from genotoxic fitness screen data: dictionary matrix, loadings matrix, and annotations. \url{https://figshare.com/articles/dataset/Webster_Supplemental_Output/14963561?file=29148153}

Table S2: Webster output from Cancer Dependency Map data: dictionary matrix, loadings matrix, and annotations. \url{https://figshare.com/articles/dataset/Webster_Supplemental_Output/14963561?file=29148159}

Table S3: Compound-to-function loadings for annotated compounds from PRISM primary and secondary screens. \url{https://figshare.com/articles/dataset/Webster_Supplemental_Output/14963561?file=29148174}

\newpage
\section{Supplemental Figures}

\begin{figure}[h]
\centering
\includegraphics[width=\textwidth]{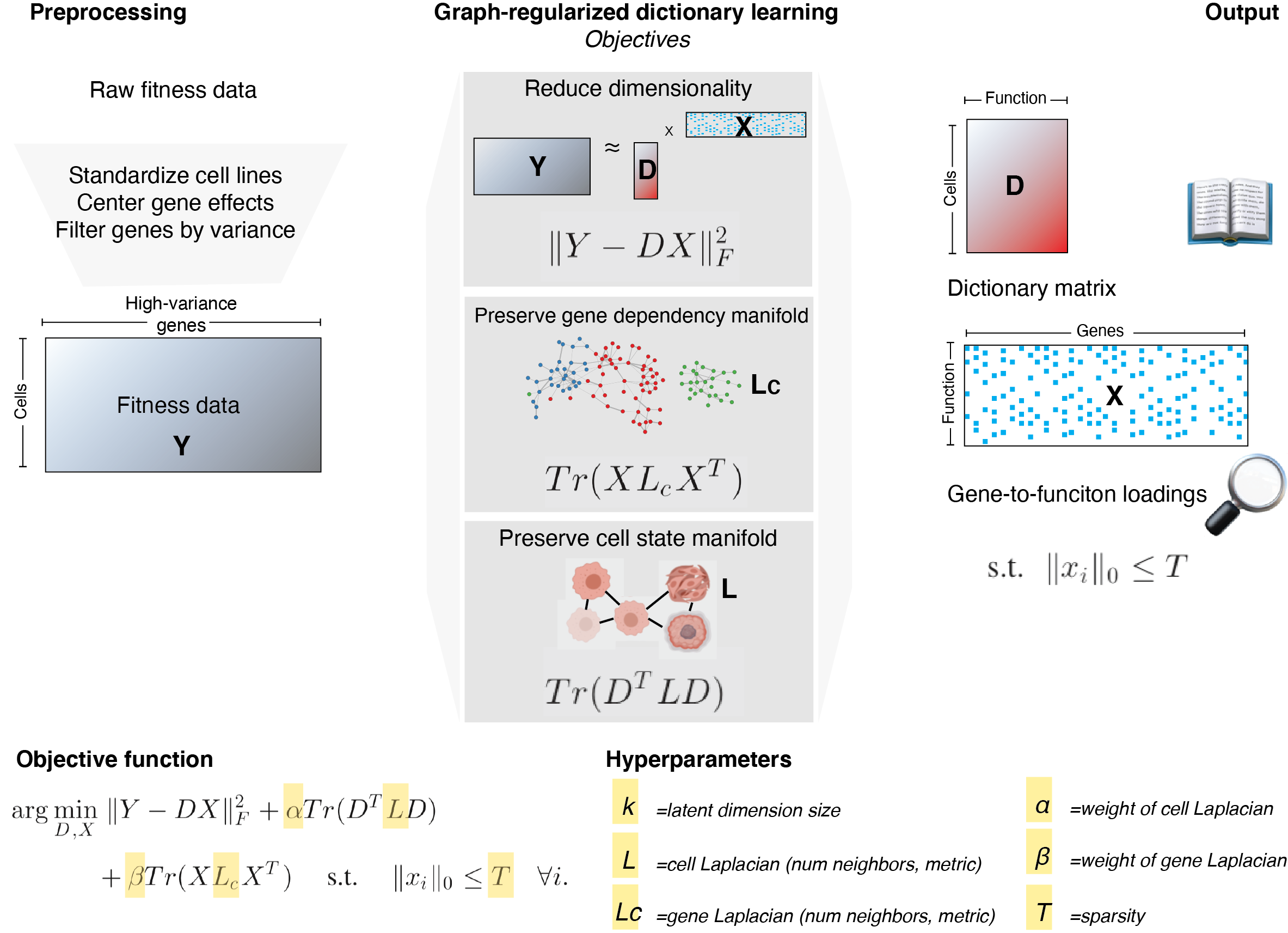}
\caption{\textit{Methodological details of Webster.} Extended version of Figure 1 showing the objective function of graph-regularized dictionary learning (Yankelevsky and Elad, 2016). Given a raw fitness dataset, Webster first preprocesses the data by standardizing cell contexts (rows), then centering gene effects (columns). It then applies a simple selection threshold to automatically choose a set of high variance gene effects (columns) to compose the input data matrix $Y$. Webster factorizes $Y$ into two low-rank matrices, $D$ and $X$, by (1) minimizing the approximation error of the low-rank factorization, (2) preserving gene effect (column) similarity from $Y$ across columns of $X$, and (3) preserving cell context (row) similarity from $Y$ across rows of $D$. Besides the key parameters $k$ and $t$, which controls the rank of the factorization and the number of non-zero entries per column of $X$, respectively, additional parameters include: the neighbor graphs used in the row and column graph-regularization (default: 5 nearest neighbors, chosen by cosine similarity); and the relative contributions of the graph regularization terms to the overall objective (default: $\alpha$ = 0.2 and $\beta$ = 0.6, as explained in (Yankelevsky and Elad, 2016)). }
\label{sf1}
\end{figure}

\begin{figure}[h]
\centering
\includegraphics[width=\textwidth]{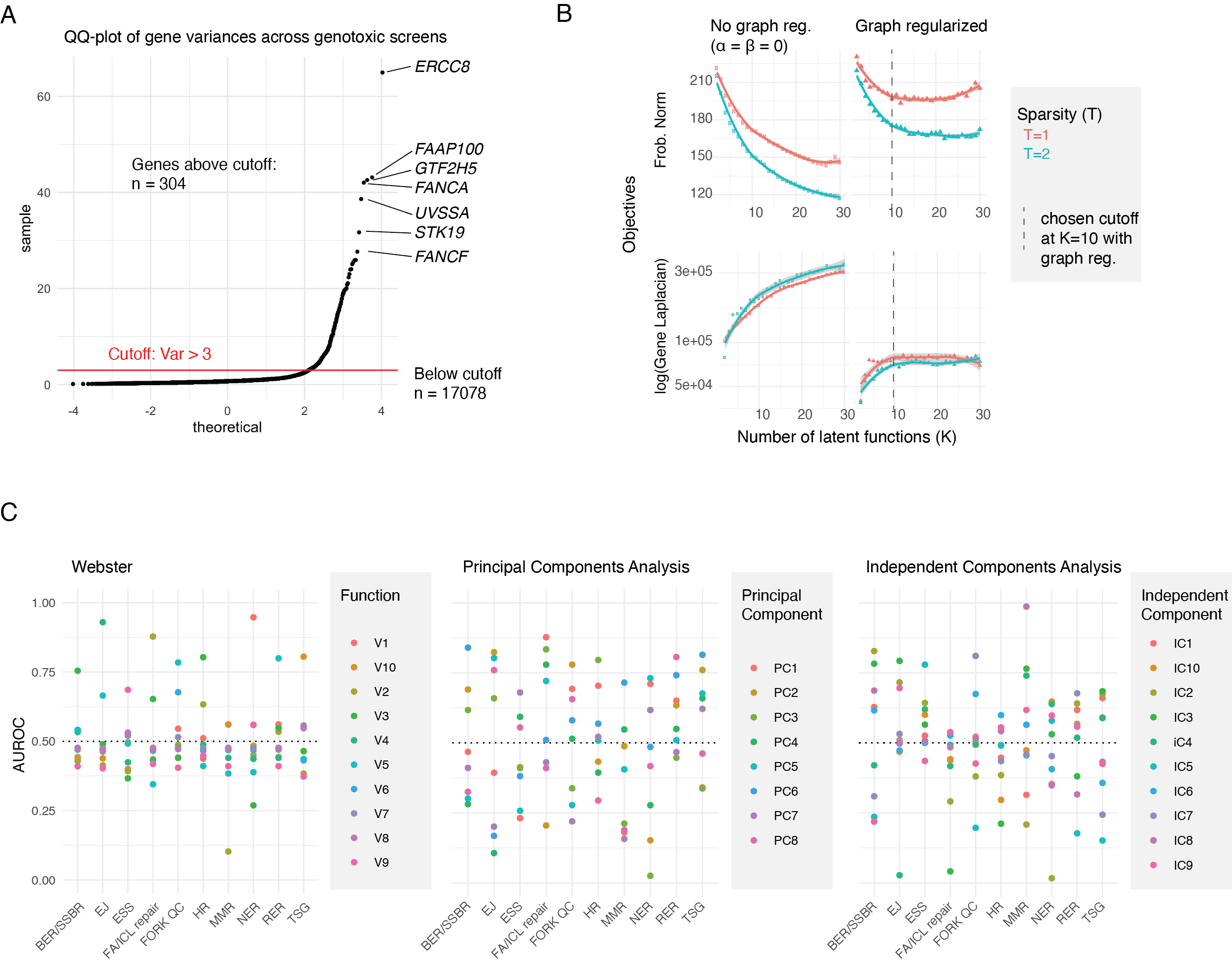}
\caption{\textit{Assessment of Webster on genotoxic fitness data. } A.	High-variance gene selection. A quantile-quantile plot is shown for the observed fitness variances for 17,382 gene effect measurements (y-axis), in comparison to a theoretical normal distribution fitted to the distribution of these variances (x-axis). A threshold is drawn to distinguish gene effects whose variance exceeds the theoretical normal distribution, resulting in 304 high-variance fitness genes chosen as input to Webster.
B.	Webster parameter grid search. Using the same data as input, we applied Webster across many values of k and t, with and without graph-regularization. Diminishing returns for both reconstruction error (Frobenius norm) and gene similarity (Gene Laplacian) are reached with k = 10 for both values of $t$. We chose $t$ = 2 in order to model pleiotropic effects in the genotoxic screening data.
C.	Interpretability of latent factors recovered from Webster, PCA and ICA. Using the literature annotations from (Olivieri et al., 2020) as ground truth, we calculated the Area Under the Receiver Operating Characteristic curve (AUROC) for each of ten genesets across each of the learned factors from all three models. The number of dictionary elements for Webster were chosen as described above; the number of PCA components was chosen with a standard elbow blot over PCA eigenvalues; the number ICA components was chosen according to (Kairov et al., 2017). The loadings for each gene over each component were used as the predictors for the AUROC metric. An AUROC $\geq$ 0.5 score indicates that positive loadings were predictive of the geneset, while an AUROC score $\geq$ 0.5 indicates that negative loadings were predictive of the geneset. 
}
\label{sf2}
\end{figure}

\begin{figure}[h]
\centering
\includegraphics[width=\textwidth]{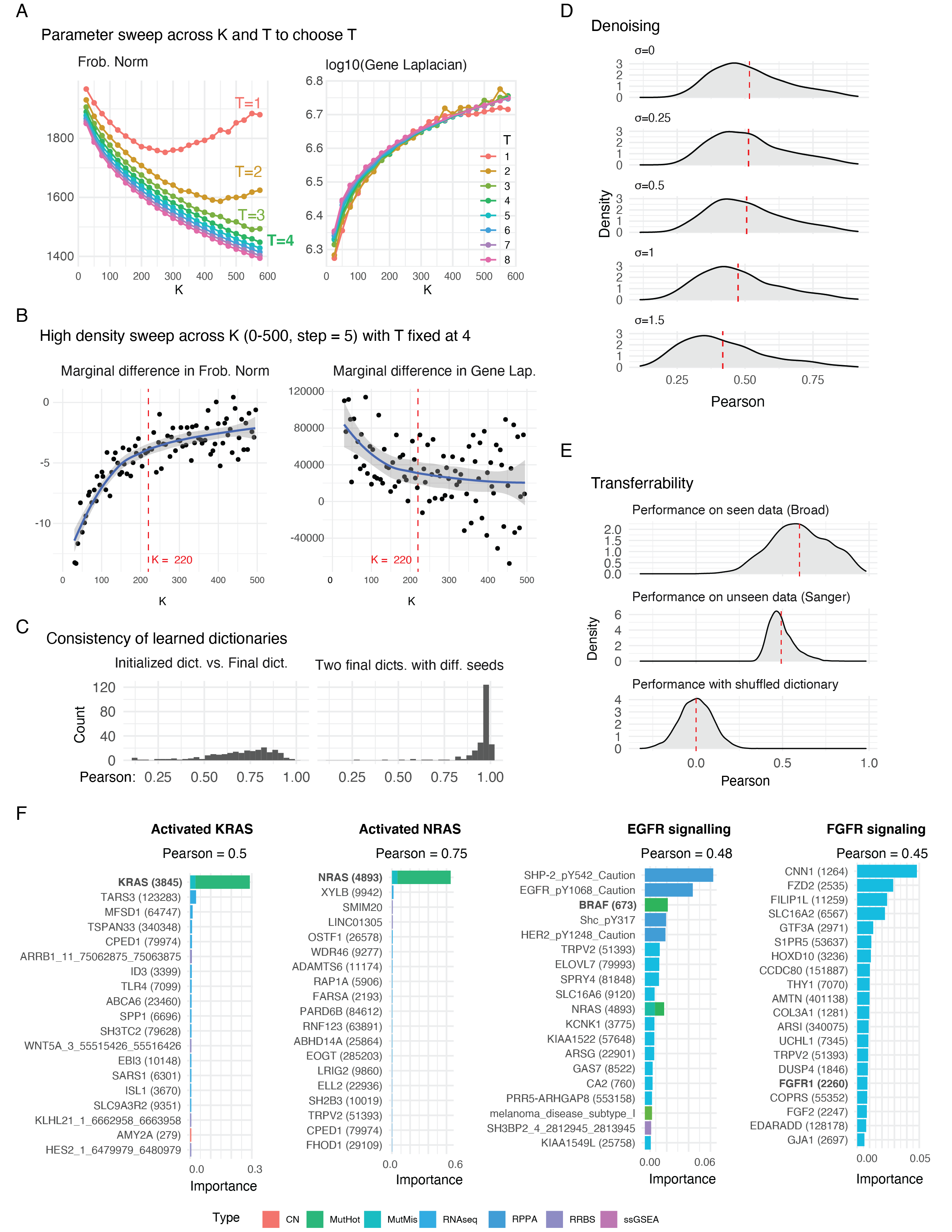}
\end{figure}

\begin{figure} []
  \caption{(Previous page.) \textit{Assessment of Webster on cancer cell fitness data. }\\A.	Webster parameter grid search. Using the cancer cell fitness data as input, we applied Webster across many values of $k$ (25 to 600, with step size 25) and $t$ (1 to 10). As $t$ = 1:3 performed poorly at large values of $k$, we chose $t$ = 4 for the factorization.\\
B.	Higher-density grid search. With $t$ = 4 fixed, we swept across $k$ (25 to 600, with step size 5) with multiple random initializations with different seeds. Plotted are the marginal improvements seen in model objectives with each additional step size of $k$, averaged over random initializations. Diminishing returns in both objectives are observed around $k$ = 220, which was chosen for the final factorization.\\
C.	Dictionary learning metrics. Left: Initialized vs. final dictionaries. In our Webster implementation, we initialize dictionary learning using a dictionary of $k$ initial gene effects chosen by $k$-medoids. Each column of the initial dictionary ($k$-medoids) was correlated to the corresponding column in the final dictionary after 20 algorithm iterations ($k$ = 220, $t$ = 4). The resulting 220 Pearson correlation values are shown as a histogram. Right: Using the same $k$-medoids dictionary as a starting point, dictionary learning was performed using two different random seed initializations. The Pearson correlations of the corresponding columns from each dictionary are shown as a histogram.\\
D.	Denoising properties of Webster. The starting fitness data was corrupted with different amounts of random noise. After splitting genes into training and test sets (3:1 split), we then applied Webster ($k$=220, $t$= 4) to learn a dictionary from the noisy training data. From this dictionary, we performed orthogonal matching pursuit to model the noisy test genes in terms of dictionary elements. We compared this reconstructed profile against the ground truth test gene profiles, which were unseen during model training. The Pearson correlation of the reconstructed test genes versus their ground truths are plotted as a distribution per noise level. The uppermost distribution ($\sigma$ = 0) corresponds to Webster’s performance in the absence of noise. Dashed red lines mark the mean of each distribution.\\
E.	Transferability of Webster dictionary elements to unseen data. Parallel genome-scale screens were performed at Broad and Sanger Institutes for 150+ common cancer cell lines, using different CRISPR-Cas9 reagents and culturing strategies. We assessed the transferability of a Webster dictionary trained on Broad data (which used the Avana CRISPR-Cas9 guide library) to model gene effects captured by the Sanger Institute (which used the Sanger CRISPR-Cas9 guide library). We learned a Webster dictionary ($k$=220, $t$=4) over the 675 cell lines screened by the Broad. We then subsetted the learned dictionary to a set of 150+ common cell lines, and used this smaller dictionary to model gene effects measured by Broad Institute or the Sanger Institute. The Pearson correlation of the reconstructed genes are plotted as a distribution. As a null comparison, we shuffled the rows of the dictionary and performed the same modeling using this shuffled dictionary. Dashed red lines mark the mean of each distribution.\\
F.	Biomarker analysis for SHOC2 functional effects. We performed a random forest regression on the fitness effect of each of the four underlying functions, using baseline -omics measurements across cancer cell lines as features (including RNA-seq bulk transcriptomic data, mutational hotspot data, protein abundance data, etc). The model performances (Pearson correlation) are shown next to barplots displaying the feature importances in the final models. Relevant biomarkers for each function are bolded. (Abbreviations; CN = copy number; MutHot = mutational hotspot; MusMis = missense mutation; RPPA = Reverse Phase Protein Array; RRBS = Reduced-representation bisulfite sequencing; ssGSEA = single sample gene set enrichment analysis)}%missing
\label{sf3}

\end{figure}

\begin{figure}[h]
\centering
\includegraphics[width=\textwidth]{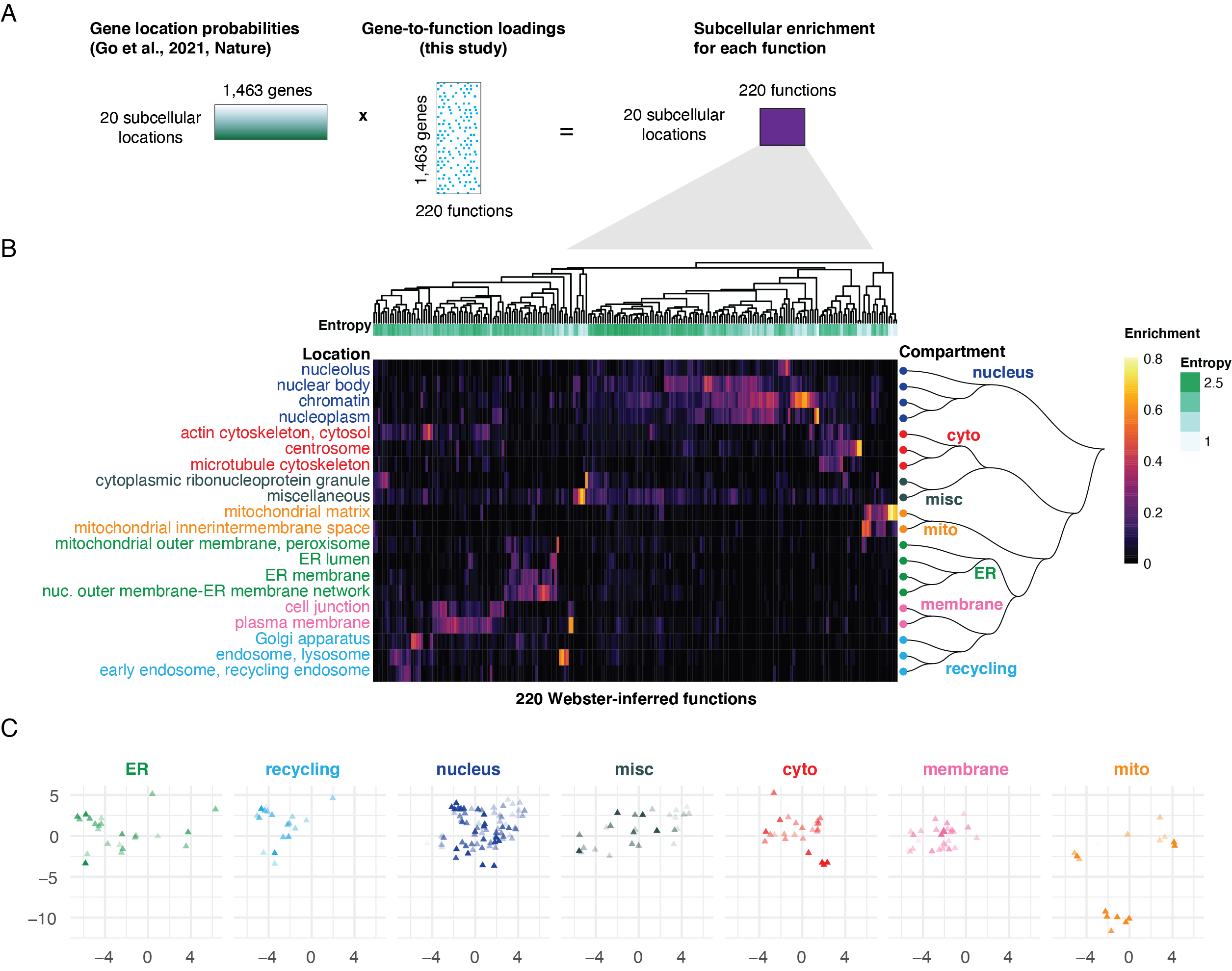}
\caption{\textit{Subcellular localization analysis from cancer fitness data. } A.	Schematic overview of subcellular localization analysis. A set of 1,463 fitness genes were also profiled in a recent subcellular localization experiment (Go et al., 2021), which reports the localization probability of each gene product over 20 inferred subcellular locations. We performed a matrix multiplication between their localization probabilities and our fitness-inferred gene-to-function loadings. The resulting matrix of 220 functions x 20 locations represents the overall distribution of localization probabilities over the learned Webster functions.
B.	A heatmap of the matrix described in A. Both rows (locations) and columns (functions) are hierarchically clustered. Clustering rows results in seven hierarchically defined cell compartments: nucleus, mitochondria, endoplasmic reticulum (ER), recycling, membrane, cytoplasm and miscellaneous. The miscellaneous category is carried over from (Go et al., 2021). Because proximity labelling proteomics were used to define subcellular locations in that study, proteins that are part of large complexes were predominantly co-labeled with other protein complex subunits, thereby decreasing their ability to infer unique subcellular locations for these proteins.
C.	Facet plot of Figure 3E, in which only functions are plotted as data points in the embedding. Functions enriched for each of the seven compartments are plotted separately.
}
\label{sf4}
\end{figure}

\begin{figure}[h]
\centering
\includegraphics[width=0.9\textwidth]{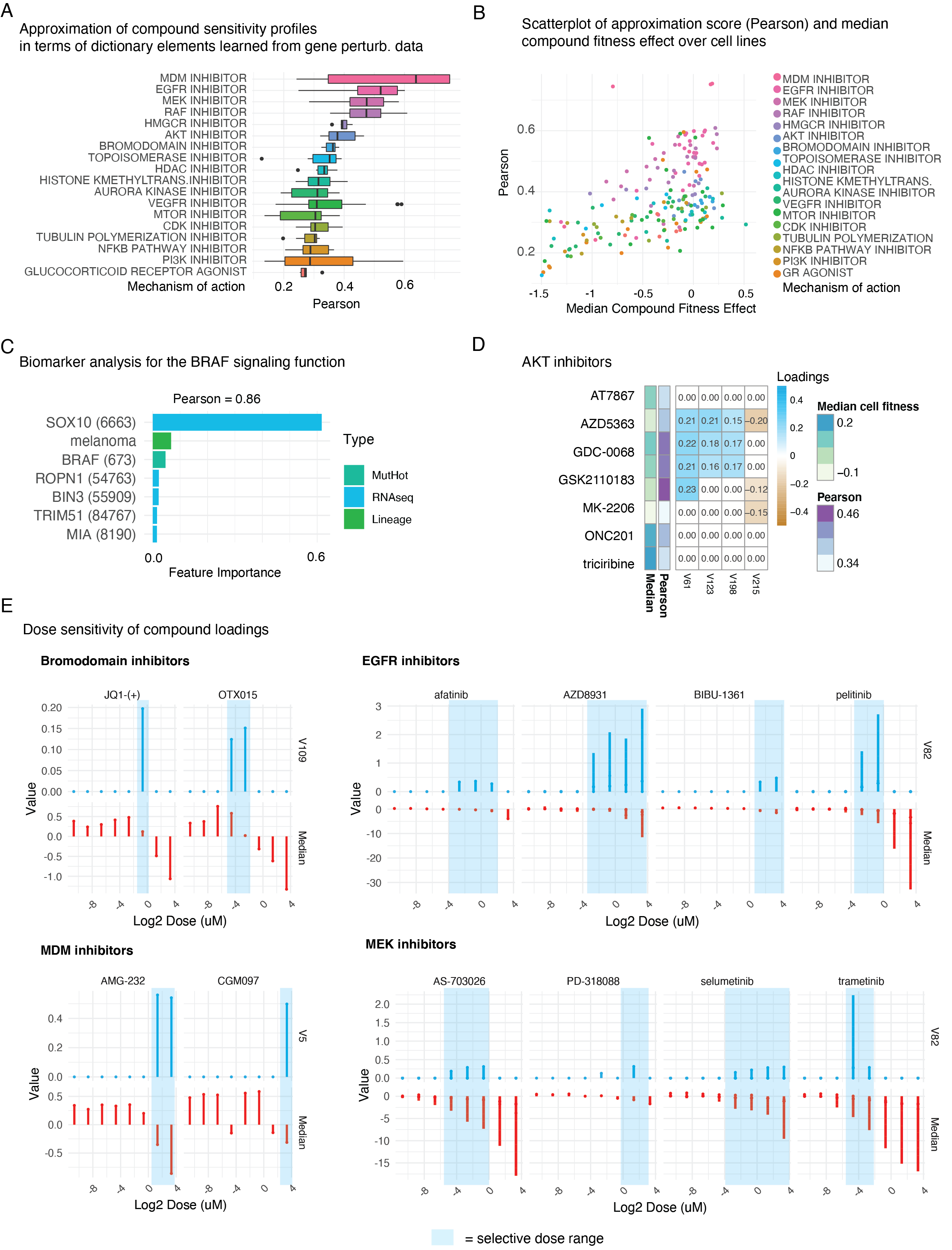}
\caption{\textit{ Compound embedding results. } A.	Compound sensitivity profiles over 360+ cancer cell lines were obtained from the PRISM Drug Repurposing dataset (Corsello et al., 2020). Each of these profiles was modeled as a sparse linear combination of four dictionary elements, using a dictionary trained on gene perturbation data (from Figure 3B). The quality of these approximations was assessed using a Pearson correlation to the original compound sensitivity profile. For each compound class, the distribution of Pearson correlations across individual drugs belonging to that class are shown in a box and whisker plot. Compound classes are ordered by their mean Pearson correlation.
B.	Each data point in the scatter plot represents one of the compounds from PRISM that was modeled in terms of gene functions. The X axis charts the median cell fitness of each compound, and the Y axis charts the Pearson correlation of the approximated profile to the measured profile.
C.	Same as Figure S3F, but for the BRAF Signaling function. 
D.	A heatmap of compound-to-function loadings. Each row represents a compound sensitivity profile for an AKT inhibitor from the PRISM primary screen (2.5 uM dose), and each column represents a Webster function learned from genetic data. Loadings values are displayed in each cell of the heatmap.
E.	Additional dose-sensitive loading plots accompanying Figure 4E.
}
\label{sf5}
\end{figure}

\end{document}